\documentclass{egpubl}

\JournalSubmission    % uncomment for submission to Computer Graphics Forum
 \electronicVersion % can be used both for the printed and electronic version

\ifpdf \usepackage[pdftex]{graphicx} \pdfcompresslevel=9
\else \usepackage[dvips]{graphicx} \fi

\PrintedOrElectronic

\usepackage{t1enc,dfadobe}

\usepackage{egweblnk}
\usepackage{cite}

\usepackage[utf8]{inputenc}
\usepackage[OT4]{fontenc}
\usepackage{amsmath}
\usepackage{amsfonts}
\usepackage{amssymb}
\usepackage{algorithm2e} 
\usepackage{stmaryrd}
\usepackage{graphicx}
\usepackage{subfigure}
\usepackage{rotating}
\usepackage{threeparttable}
\usepackage{mathtools}

\newcommand{\tensor}[1]{\mathcal{#1}}
\newcommand{\M}[1]{\mathbf{#1}}
\newcommand{\tucker}[2]{\llbracket#1;#2\rrbracket}
\newcommand{\set}[1]{\mathfrak{#1}}

\newcommand{\vecb}[1]{\mathbf{#1}}
\newcommand{\R}{\mathbb{R}}
\newcommand{\dd}{\mathrm{d}}

\title{Compression of animated 3D models using HO-SVD}
\author{Michał Romaszewski, Piotr Gawron, Sebastian Opozda}

\begin{document}

\maketitle

\begin{abstract}
This work presents an analysis of Higher Order Singular Value
Decomposition (HO-SVD) applied to lossy compression of 3D mesh animations. We
describe strategies for choosing a number of preserved spatial and temporal
components after tensor decomposition. Compression error is measured using
three metrics (MSE, Hausdorff, MSDM). Results are compared with a
method based on Principal Component Analysis (PCA) and presented on a set of
animations with typical mesh deformations.
\end{abstract}

\section{Introduction}
The goal of this paper is to provide an analysis of Higher Order
Singular Value Decomposition \cite{lathauwer2000multilinear} (HO-SVD) applied
to lossy compression of 3D mesh animations. The paper includes an estimation of
compression quality using three error metrics, a discussion about selection
of parameters and a comparison with a method based on Principal Component
Analysis (PCA). We provide a simple, heuristic method for HO-SVD parameter
selection. Results are presented using a
diverse set of well-known mesh animations, representing several common cases of
3D shape deformation.

HO-SVD is a multi-linear generalization of Singular Value Decomposition. It~has
been shown (e.g. in \cite{Inoue:2005}) that HO-SVD is an efficient method for
dimensionality reduction of data represented as tensors, also called N-way
arrays. While consecutive frames of a 3D animation can naturally be represented
as a 3-mode tensor (a data cube), by stacking arrays of their vertices,
application of HO-SVD for such data requires additional considerations.

When using PCA-based compression, only a single parameter related to the number
of preserved principal components is needed. Usually (e.g. \cite{AM2000},
\cite{Sattler:2005}), dimensionality reduction is applied to animation
frames, reducing their number to a sequence of significant key-frames. On the  
contrary, HO-SVD-based compression allows for multidimensional reduction of the
data tensor, so it is possible to achieve the desired compression ratio with
multiple sets of parameters. In our experiment we truncated the number of
mode-$1$ (mesh vertices) and mode-$3$ (animation frames) components obtained
through tensor decomposition. Reduction of mode-$2$ components (3D coordinates
of vertices) is not advised since it results in a significant loss of
information and low quality of reconstructed data. We assume that the best set
of parameters for a desired compression ratio is the one introducing the lowest
distortion of reconstructed objects.  

\begin{figure*}[h]
\centering
	\subfigure[]{\includegraphics[width=0.24\textwidth]{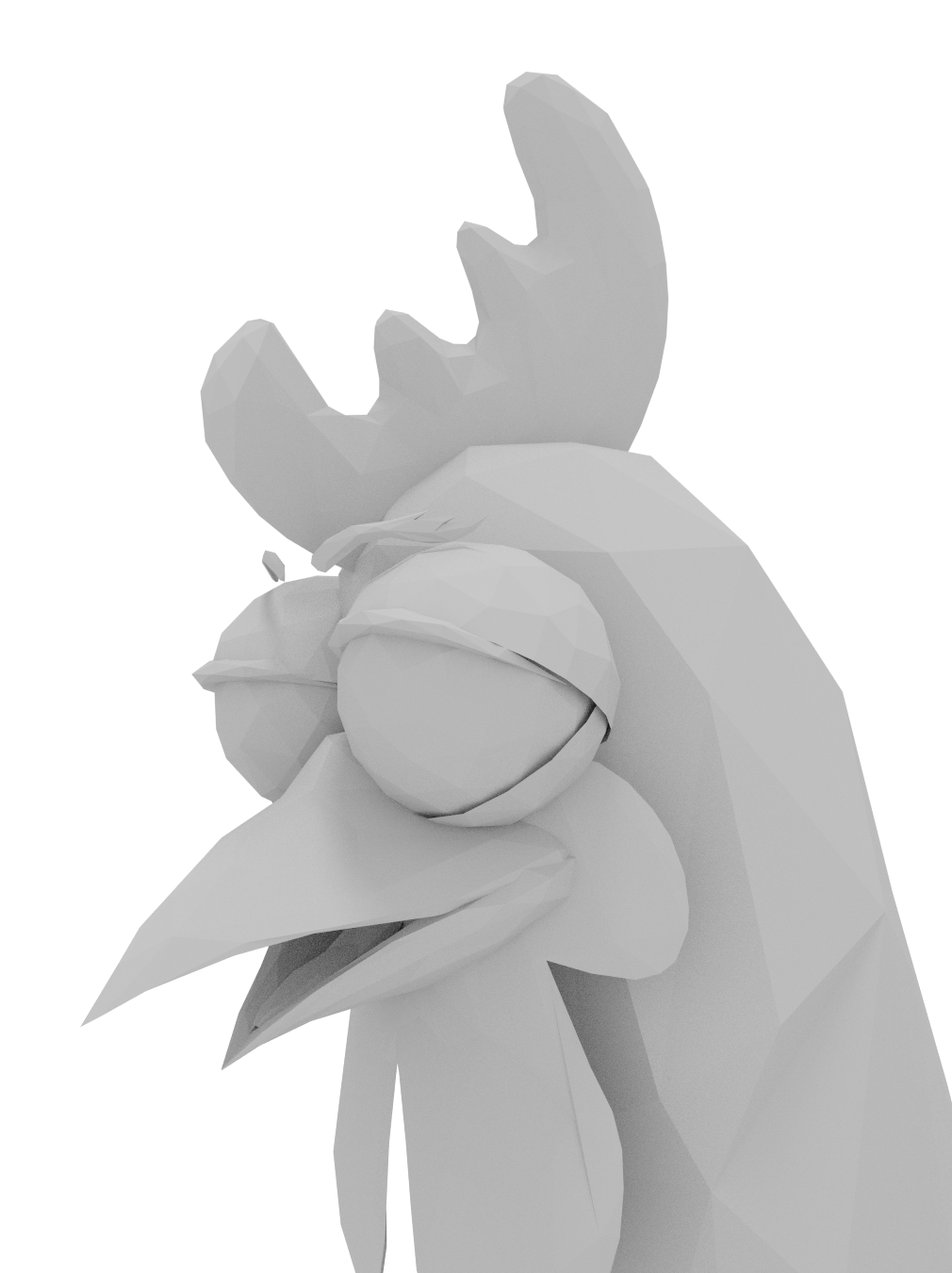}}
	\subfigure[]{\includegraphics[width=0.24\textwidth]{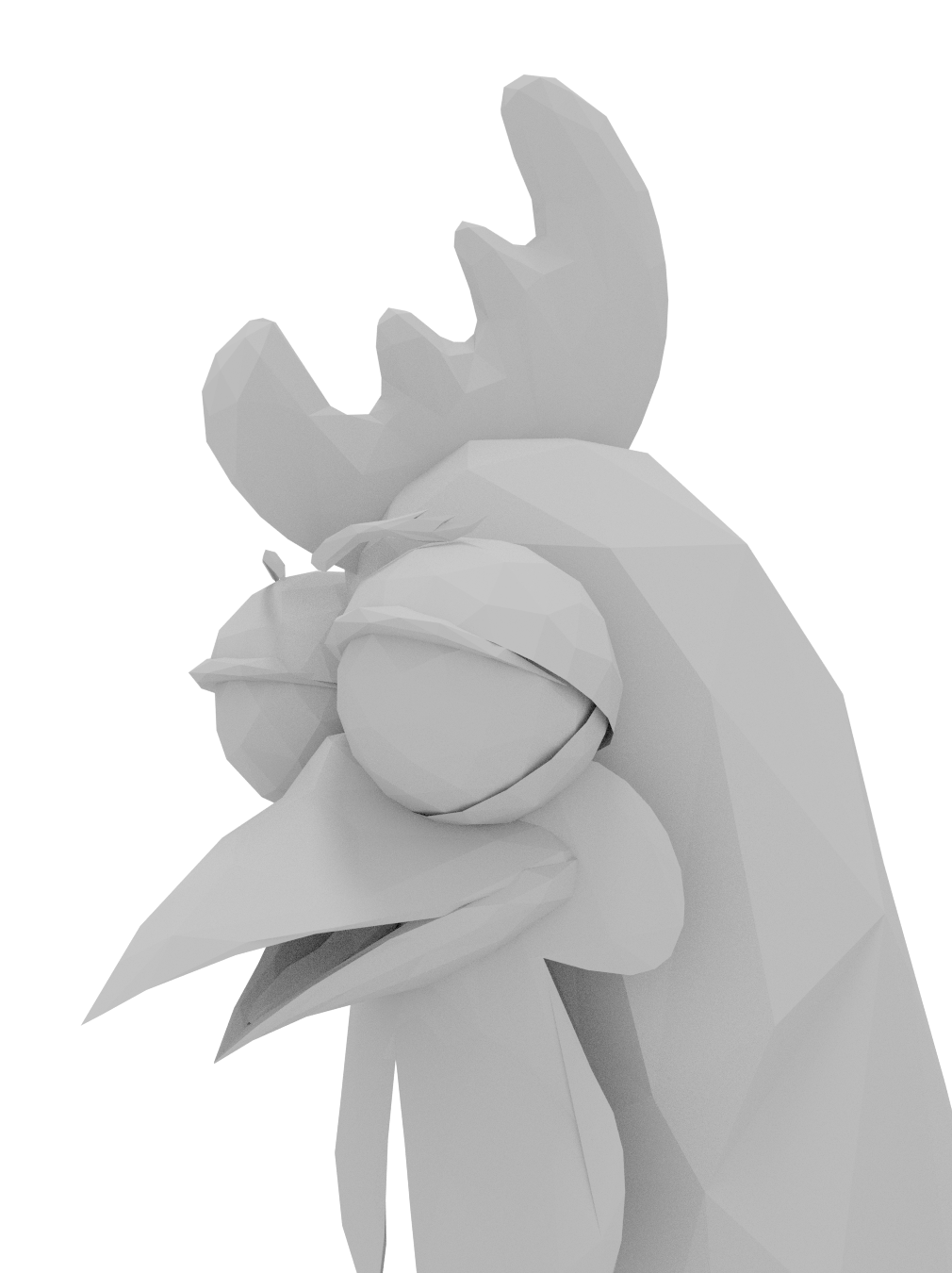}}
	\subfigure[]{\includegraphics[width=0.24\textwidth]{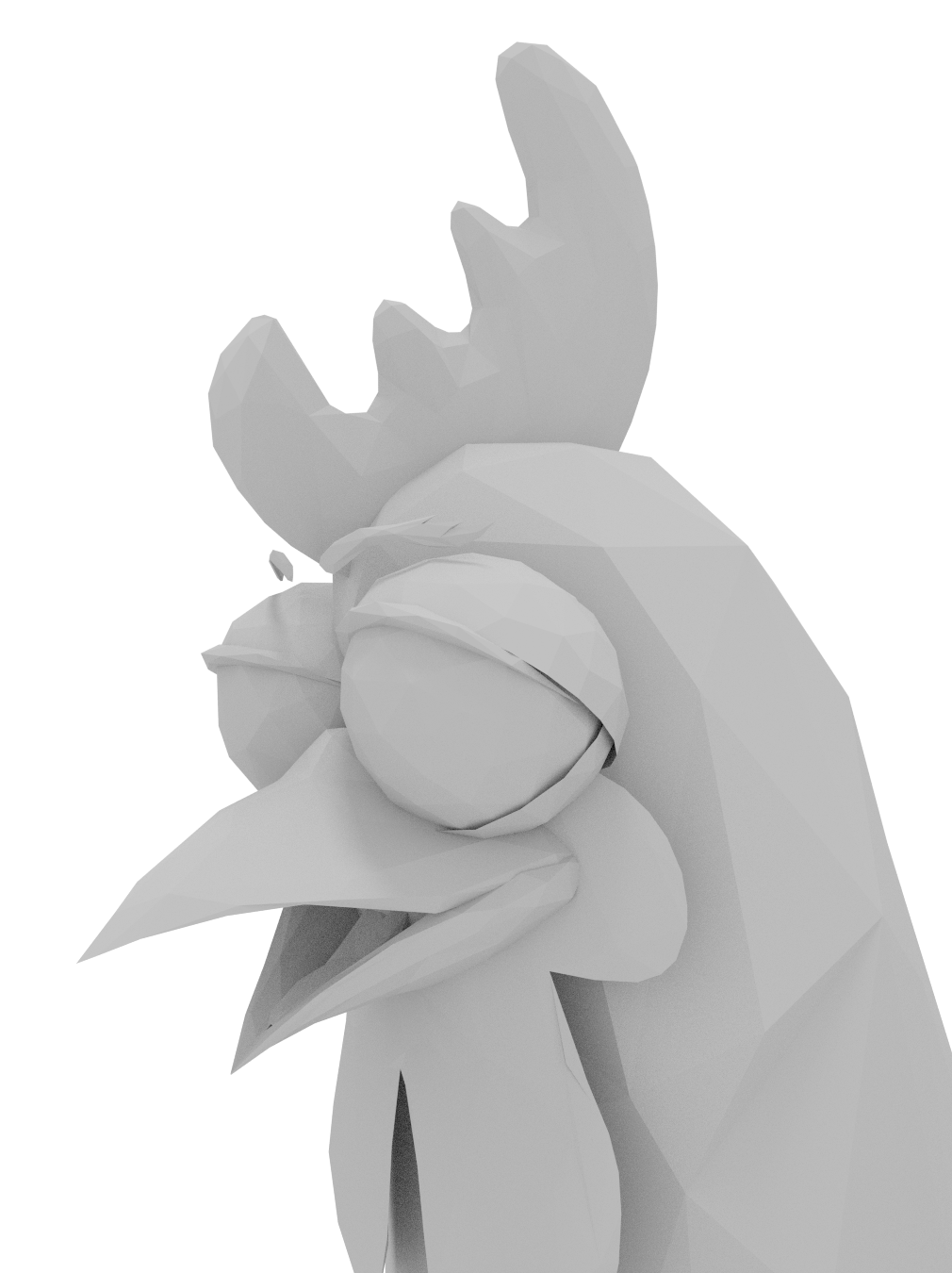}}
	\subfigure[]{\includegraphics[width=0.24\textwidth]{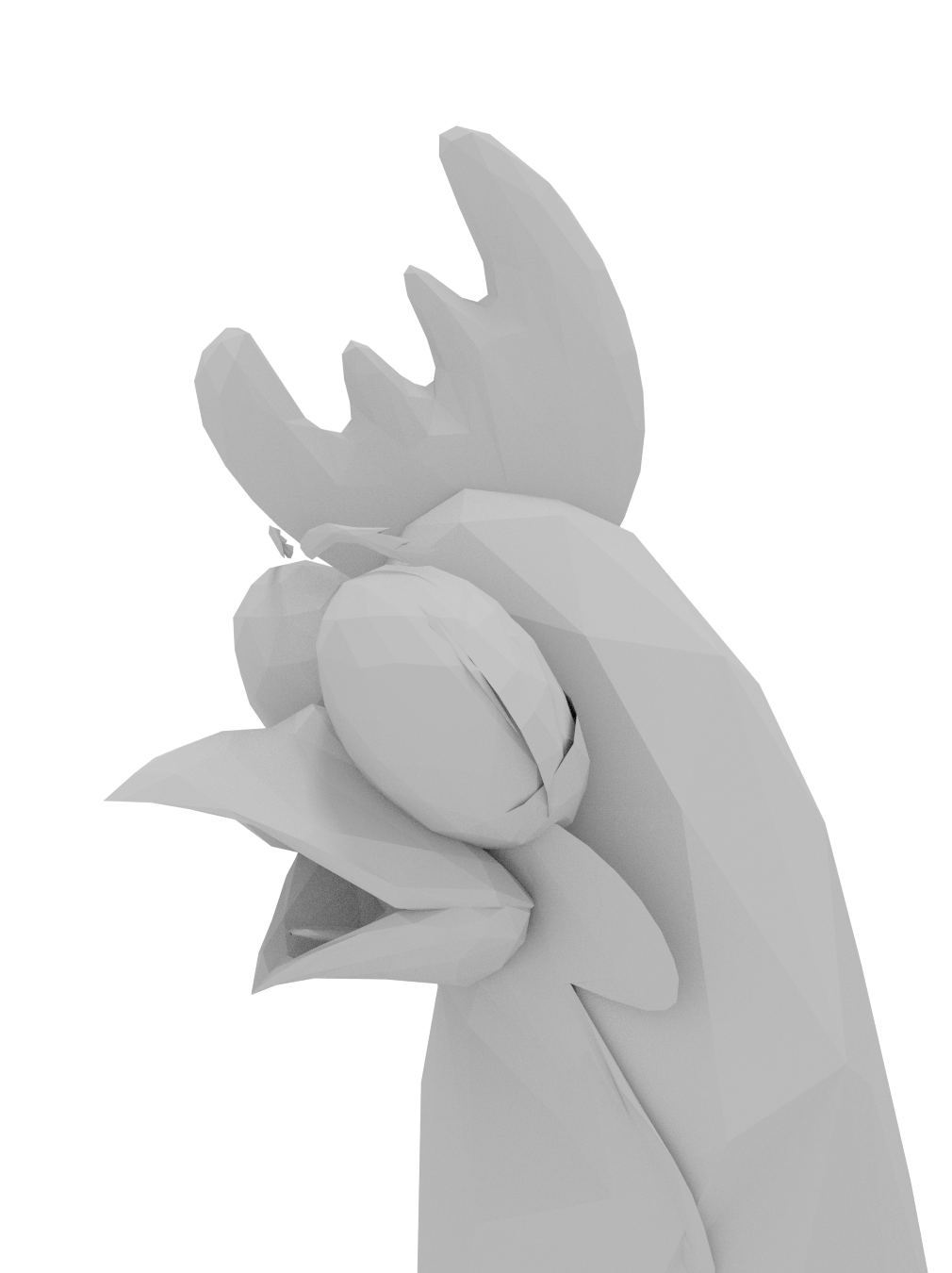}}
  \caption{A fragment of a reconstructed animation sequence for \emph{Chicken} animation. Panel (a) presents an original model, in further panels the data tensor is compressed to (b): 5.1\%, (c): 2.1\%, and (d): 1.1\% of its original size.}
  \label{fig:visualization_CC_head}
\end{figure*}

For estimation of the quality of lossy compression we used three metrics. The
Mean Squared Error (MSE) and the Hausdorff distance are both widely used for
measuring 3D mesh distortions. Additionally, we decided to include a third
metric, the Mesh Structural Distortion Measure (MSDM), since according to
\cite{Guillaume:2006}, it correlates well with human perception of errors in 3D
data. Fig.\ref{fig:visualization_CC_head} presents an example of a distortion,
resulting from a reconstruction of an animation compressed with HO-SVD.

The article is organised as follows. In the two following subsections, the
related work and HO-SVD decomposition are presented. Definitions and
methodology of our experiments are presented in Section~\ref{sec:method}.
Obtained results can be found in Section~\ref{sec:results}, while their summary
along with our comments are presented in Section~\ref{sec:conclusions}.

\subsection{Related work}
\label{sub:related_work}
Due to their amount, data generated by using 3D scanners or animation software
require effective compression methods for their storage, transmission, and
processing. An in-depth survey of 3D mesh compression techniques along with
formulation of related problems is provided in \cite{Peng:2005}. Various
approaches for 3D data modelling, e.g. using efficient octree coding and
attribute quantization \cite{Huang:2008:genericpccoding} or combining surface
and probabilistic approximation \cite{przemg:2013:smm} are used to effectively
store and reproduce digitized objects.

Frame-based Animated Mesh Compression was recently promoted within the MPEG-4
standard as Amendement 2 to part 16 (AFX, Animation Framework eXtension) and is
described in \cite{MZP2008}. A representation of animation sequences using
Principal Component Analysis (PCA) was introduced in \cite{AM2000}. This
approach was further refined in \cite{Sattler:2005} where authors performed
motion clustering on an animation and applied PCA to its subsegments,
substantially improving the reconstruction quality. A summary of PCA-based
applications to 3D animation compression, along with a broad analysis of
related topics is provided in \cite{Amjoun:2009}.

Higher Order Singular Value Decomposition (HO-SVD) may be treated as a natural
extension of PCA for high-dimensional data. A~survey of tensor properties as
well as the description of higher-order tensor decomposition is provided in
\cite{TensorReview2009}. An application of HO-SVD to face recognition is
presented in \cite{Hongcheng:2003}. The authors used a data tensor consisting
of a classification index, along with subspaces corresponding to human facial
features and expressions. After tensor decomposition, face recognition was
performed using the $k$-NN classifier.

An interesting representation of 3D animation using HO-SVD is provided in
\cite{Akhter:2012}, with applications including dimensionality reduction,
denoising and gap filling.  HO-SVD was also used in \cite{Mukai:2007} for
level-of-detail reduction in animations of human crowds, where encoded motion
tensors are represented with a number of components dependent on the distance
from the camera.

\subsection{Higher Order Singular Value Decomposition}
\label{sec:hosvd}
Higher Order Singular Value Decomposition, also called Tucker decomposition, is
a~generalisation of SVD from matrices to tensors (N-way arrays). In this
section we recall basic facts about tensors and HO-SVD. We follow conventions
presented in \cite{TensorReview2009}.

To describe this decomposition, first we will recall basic notions regarding
operations on tensors. Let a tensor 
\begin{equation}\label{equ:tensor}
\tensor{T}=
\{t_{i_1,i_2,\ldots,i_n}\}_{i_1,i_2,\ldots,i_n=0}^{I_1-1,I_2-1,\ldots,I_N-1}
\in \R^{I_1,I_2,\ldots,I_N} 
\end{equation} 
be given --- we say that this tensor
has $n$ modes. Each of the indices corresponds to one of the modes
\textit{i.e.} $i_l$ to mode $l$.

By \emph{multiplication} of tensor $\tensor{T}$ by matrix 
$\M{U}=\{u_{i_l d}\}_{i_l,d=0}^{I_l-1,D} \in \R^{I_l, D}$ 
in mode $l$ we define tensor  $\tensor{T}'\in 
\R^{I_1,\ldots,I_{l-1},D,I_{l+1},\ldots,I_N}$, such that:
\begin{equation}
\tensor{T}'=(\tensor{T}\times_l \M{U})_
{i_1\ldots i_{l-1} d\, i_{l+1}\ldots i_N}=\sum_{{i}_l=0}^{I_l-1} 
t_{i_1 i_2\ldots{i}_l\ldots i_N} u_{i_l d}.
\end{equation}

By \emph{unfolding} tensor $\tensor{T}$ in mode $l$ we define matrix 
$\M{T}_{(l)}$ such that
\begin{equation}
\label{equation:unfolding}
(\M{T}_{(l)})_{i,j}=t_{i_1\ldots i_{l-1} j\, i_{l+1}\ldots 
i_N},
\end{equation}
where \[i=1+\sum_{\substack{k=1 \\ l\neq l}}^{N}J_k\text{ \ and \ }
J_k=\prod_{\substack{m=1 \\ m\neq l}}^{k-1}I_m.\]

Given tensor $\tensor{T}$, defined as in Eq.~(\ref{equ:tensor}), a new \emph{sub-tensor}
$\tensor{T}_{{i}_n=\alpha}$ can be created according to the equation with the
following elements:
\begin{equation}\label{equ:subtensor}
\tensor{T}_{{i}_l=\alpha}=
\{t_{i_1 i_2 \ldots i_{l-1} i_{l+1} \ldots i_n}\}_
{i_1=0,i_2=0,\ldots,i_l=\alpha,\ldots,i_n=0}^
{I_1-1,I_2-1,\ldots,\alpha,\ldots,I_N-1} \in \R^{I_1,I_2,\ldots, 1,\ldots,I_N}.
\end{equation}

The \emph{scalar product} $\langle \tensor{A}, \tensor{B} \rangle$ of tensors 
$\tensor{A}, \tensor{B}\in\R^{I_1,I_2,\ldots,I_N}$ is defined as
\begin{equation}
\langle \tensor{A}, \tensor{B} \rangle=
\sum\limits_{i_1=0}^{I_1-1}
\sum\limits_{i_2=0}^{I_2-1}
\ldots
\sum\limits_{i_N=0}^{I_N-1}
b_{i_1,i_2,\ldots,i_n} a_{i_1,i_2,\ldots,i_n}.
\end{equation}
We say that if scalar product of tensors equals 0, then they are orthogonal.

The \emph{Frobenius norm} of tensor $\tensor{T}$ is given by
\begin{equation}
||\tensor{T}||=\sqrt{\langle \tensor{T}, \tensor{T} \rangle}.
\end{equation}

Given tensor $\tensor{T}$, in order to find its HO-SVD, in
the form of the so called Tucker operator
$\tucker{\tensor{C}}{\M{U}^{(1)},\ldots,\M{U}^{(N)}}$,
such that $\tensor{C}\in \R^{I_1,\ldots,I_N}$ and $\M{U}^{(k)}\in \R^{I_k \times
I_k}$ are orthogonal matrices, Algorithm~\ref{alg:hosvd} can be used.

\begin{algorithm}[h]
\KwIn{Data Tensor $\tensor{T}$}
\KwOut{Tucker operator $\tucker{C}{\M{U}^{(1)},\ldots,\M{U}^{(N)}}$}
\For{$k \in \{1,\ldots,N\}$}{
	$\M{U}^{(k)} = \text{ left singular vectors of } T_{(k)} \text{ in unfolding } 
k$\;
	}
	$C=\tensor{T} \times_1 \M{U}^{(1)T} \times_2 \M{U}^{(2)T} \ldots \times_N 
	\M{U}^{(N)T}$\;
	\Return{$\tucker{C}{\M{U}^{(1)},\ldots,\M{U}^{(N)}}$}\;
\caption{HO-SVD algorithm}
\label{alg:hosvd}
\end{algorithm}

Tensor $\tensor{C}$ is called the core tensor and has the following
useful properties.
\begin{itemize}
\item Reconstruction:
\begin{equation}
\tensor{T}=\tensor{C} \times_1 \M{U}^{(1)} \times_2 \M{U}^{(2)} \times_{3} \ldots
\times_N \M{U}^{(N)}, 
\end{equation}
where $\M{U}^{(i)}$ are orthogonal matrices;

\item Orthogonality:
\begin{equation}
\langle\tensor{C}_{{i}_l=\alpha},\tensor{C}_{{i}_l=\beta}\rangle=0
\end{equation}
for all possible values of $l$, $\alpha$ and $\beta$, such that $\alpha \neq \beta$;

\item Order of sub-tensor norms:
\begin{equation}
||\tensor{C}_{{i}_n=1}||\leq ||\tensor{C}_{{i}_n=2}|| \leq \ldots \leq ||\tensor{C}_{{i}_n=I_n}||
\end{equation}
for all $n$.
\end{itemize}

Therefore, informally, one can say that larger values of a core tensor are
denoted by low values of indices. This property is the basis for the
development of compression algorithms based on HO-SVD.

\begin{figure*}[!h]
\centering
\includegraphics{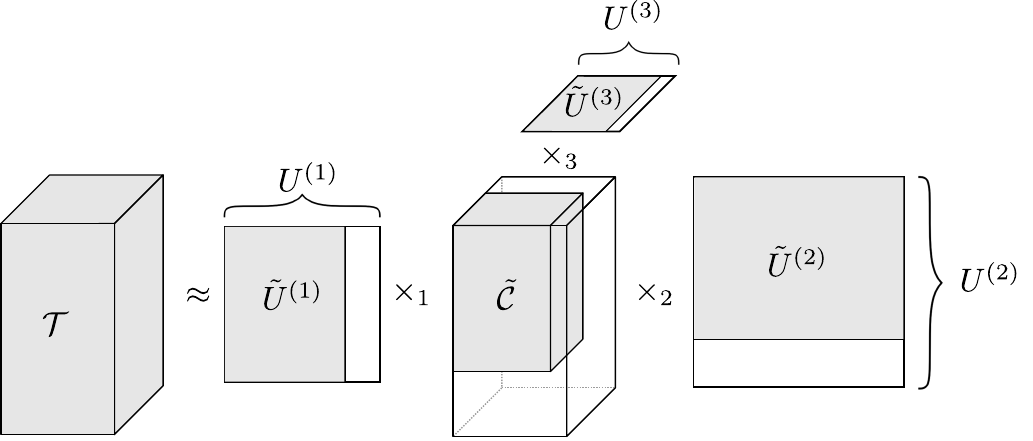}
\caption{Truncated HO-SVD decomposition of tensor $\tensor{T}$. Its
approximation, tensor $\tilde{\tensor{T}}$, can be reconstructed from a truncated
tucker operator
$\tucker{\tilde{C}}{{\tilde{\M{U}}^{(1)},\tilde{\M{U}}^{(2)},\tilde{\M{U}}^{(3)}}}$. 
The
visualization is inspired by \cite{TensorReview2009}.}
\label{fig:tucker_decompostion}
\end{figure*}

Formally 
\begin{equation}
\tensor{\tilde{T}}=\tilde{\tensor{C}} \times_1 \tilde{\M{U}}^{(1)} \times_2 \tilde{\M{U}}^{(2)} \times_{3} \ldots
\times_N \tilde{\M{U}}^{(N)}, 
\end{equation}
where 
\begin{equation}
\tensor{\tilde{C}}=
\{c_{i_1,i_2,\ldots,i_n}\}_{i_1,i_2,\ldots,i_n=0}^{R_1-1,R_2-1,\ldots,R_N-1} \in \R^{R_1,R_2,\ldots,R_N}
\end{equation}
is a truncated tensor in such a way that in each mode $l$ indices span from 0
to $R_l-1\leq I_l-1$ and $\tilde{\M{U}}^{(l)}\in \R^{R_l\times I_l}$ matrices
whose columns are orthonormal and rows form orthonormal basis in respective
vector spaces. A visualization of $3$-mode truncated tensor is provided in
Fig.~\ref{fig:tucker_decompostion}

Given $(R_l)_{l=1}^{N}$ one can form tensor $\tensor{\tilde{T}}$ that 
approximates tensor $\tensor{T}$ in the sense of their euclidean distance
$||\tensor{\tilde{T}}-\tensor{T}||$. This approximation can be exploited to
form lossy compression algorithms of signals that are indexed by more than two 
indices. It should by noted that the choice of $(R_l)_{l=1}^{N}$ in a given 
application is non-obvious and depends on the properties of processed signals.

\section{Method}\label{sec:method}
Our experiments aim to assess the effectiveness of HO-SVD for compression of 3D
animations. 
Additionally we want to present a clear strategy for choosing the
proportion of preserved spatial and temporal components in order to maintain
good quality of reconstructed data. We will present results using multiple
error metrics and compare them to a method based on PCA.

\begin{figure*}[!h]
\centering
\includegraphics{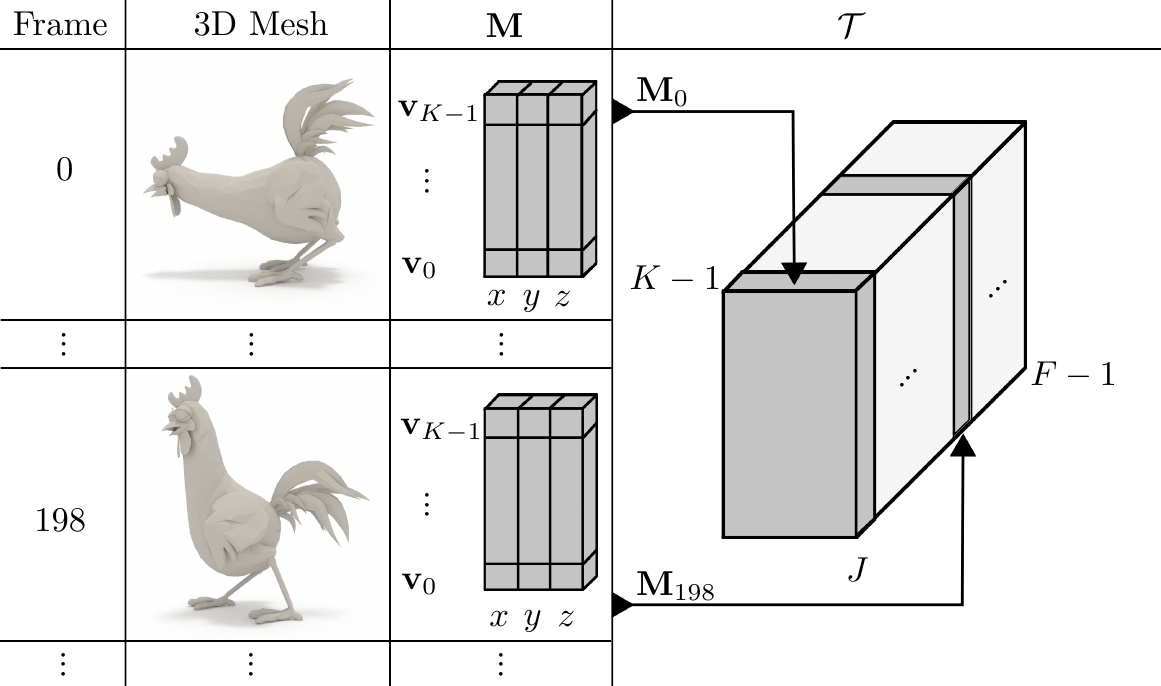}
\caption{Visualization of data tensor $\tensor{T}$. It is formed by stacking
vertices of meshes, corresponding to $F$ animation frames, represented as
$K\times J$ arrays.}
\label{fig:tensor}
\end{figure*}

\subsection{Input data}\label{sec:datamodel}
Three-dimensional mesh will be treated as a $K\times J$ matrix $\M{M}$, with
mesh vertices $\vecb{v}_i\in\R^J, i\in \{0,\dots,K-1\}$ as rows, together with
a set of edges $\set{G}$. We denote $J=3$ as a number of spatial dimensions. An
animation consists of $F$ successive frames enumerated with $k$, each
containing a mesh $\M{M}_k, k\in \{0,\dots,F-1\}$, with the same topology, but
different coordinates of vertices. Therefore, input data can form tensor
$\tensor{T}=t_{i,j,k}\in \R^{K\times J\times F}$ while meshes $\M{M}_k$,
following the notation in \cite{TensorReview2009}, form frontal slices
$\tensor{T}_{::i}$. Tensor visualization is presented in Fig.~\ref{fig:tensor}.
A~pair
$(\tensor{T},\set{G})$ contains all available information about the animation.
We apply compression only to $\tensor{T}$, a set of edges $\set{G}$ is used
only for data visualization.

\subsection{Compression testing procedure}\label{sec:datamodel2}
Algorithm \ref{alg:experiment_hosvd} presents stages of the procedure aimed at
calculating the quality of mesh reconstruction.

\begin{algorithm}[!h]
\KwIn{Data Tensor $\tensor{T}$, Compression rate $CR$, Quality metric $d$}
\KwOut{Quality of $\tensor{T}^\prime$}
\tcc{$\tensor{X}$ is a normalised tensor} 
\tcc{$\set{R}$ is a sequence of homography matrices}
$\tensor{X}$, $\set{R}$ = Rigid Motion Estimation($\tensor{T}$)\;
\tcc{$\tucker{C}{\M{U}^{(1)},\M{U}^{(2)},\M{U}^{(3)}}$ is the Tucker operator}
$\tucker{C}{\M{U}^{(1)},\textbf{U}^{(2)},\textbf{U}^{(3)}}$ = HOSVD Decomposition(X)\;
\tcc{$VTF$ is a scalar of Vertex-to-Frame ratio}
$VTF$ = Estimate VTF($\tucker{C}{\M{U}^{(1)},\M{U}^{(2)},\M{U}^{(3)}}$, $CR$)\;
$\tucker{\tilde{C}}{{\tilde{\M{U}}^{(1)},\tilde{\M{U}}^{(2)},\tilde{\M{U}}^{(3)}}}$ = 
Truncate(($\tucker{C}{{\M{U}^{(1)},\M{U}^{(2)},\M{U}^{(3)}}}$)\\
$\tilde{\tensor{T}}$ = 
Reconstruct($\tucker{\tilde{C}}{{\tilde{\M{U}}^{(1)}, \tilde{\M{U}}^{(2)},\tilde{\M{U}}^{(3)}}}$, 
            $VTF$, 
            $CR$, 
            $\set{R}$)\;
\Return{$\dd(\tensor{T},\tilde{\tensor{T}})$}\;
\caption{A procedure for estimating the quality of HO-SVD
compression for 3D animation. Similar procedure is performed for PCA.}
\label{alg:experiment_hosvd}
\end{algorithm}

The algorithm consist of the following steps:
\begin{enumerate}
\item \emph{Rigid motion estimation}:\\
The rigid motion of a mesh over the whole animation is estimated and subtracted
from consecutive frames. Sequence $\set{R}=(\M{R}_{i})_{i=0}^{F-1}$ is created
where $\M{R}_{i}$ is a matrix of an affine transformation between frame $i$ and
$0$. Consecutive frames are translated into the coordinates of the first frame,
forming normalised data tensor $\tensor{X}: \forall_i
\tensor{X}_{::i}=\tensor{T}_{::i} \M{R}_{i}^T$, used in the following steps.
\item \emph{Tensor decomposition}:\\ Tensor $\tensor{X}$ is decomposed using
HO-SVD, so Tucker decomposition\\
$\tucker{C}{{\M{U}^{(1)},\M{U}^{(2)},\M{U}^{(3)}}}$ is obtained. \item
\emph{HO-SVD compression parameter estimation}:\\ We estimate the
Vertices-to-Frame ratio $VTF$ between a number of spatial and temporal
components of the decomposed tensor $\tensor{X}$, required to achieve the
desired compression rate, by searching for the best set in the parameter space.
\item \emph{HO-SVD compression}:\\
Compression is performed by truncating a number of spatial and temporal
components of $\tucker{C}{{\M{U}^{(1)},\M{U}^{(2)},\M{U}^{(3)}}}$, forming
truncated Tucker operator
$\tucker{\tilde{C}}{{\tilde{\M{U}}^{(1)},\tilde{\M{U}}^{(2)},\tilde{\M{U}}^{(3)}}}$.
\item \emph{Reconstruction}:\\
Tensor $\tensor{\tilde{X}}$ is reconstructed from $\tucker{\tilde{C}}{{\tilde{\M{U}}^{(1)},\tilde{\M{U}}^{(2)},\tilde{\M{U}}^{(3)}}}$. Then, frames of the animation are
transformed to their original coordinates by applying a corresponding inverse
transformation from $\set{R}$ to each frame, forming tensor $\tensor{\tilde{T}}:
\forall_i \tensor{\tilde{T}}_{::i}={\tensor{\tilde{X}}}_{::i}
(\M{R}_{i}^T)^{-1}$.
\item \emph{Estimation of the reconstruction quality}:\\
The quality of reconstruction $\dd( \tensor{T},\tensor{\tilde{T}} )$ is
computed using three 3D quality metrics: Mean Squared Error, Hausdorff distance
and Mesh Structural Distortion Measure.
\end{enumerate}

Steps 1 to 4 correspond to the animation compression process. Step 5
represents data decompression. After step 4, an additional lossless
compression of floating-point data may be performed.
We provide a detailed description of these steps in the following subsections.

\subsection{Rigid motion estimation}
\label{subsection:rigid_motion_estimation}
The first step of the algorithm follows the idea from \cite{AM2000}. If a set
of edges $\set{G}$ of mesh $\M{M}$ is constant through the animation, mesh
state in frame $i$, can be described by the sum of changes applied to $\M{M}$
in each
frame: 
$\M{M}_i=\sum_{j=1}^{i}(\M{M}_j-\M{M}_{j-1})=\sum_{j=1}^{i}\Delta{\M{M}_j}$ . 
Assuming that animation is represented in homogeneous coordinates, the
difference between two consecutive frames $\Delta{\M{M}_j}= \M{D}_j
\M{R}_j^{T}$, where $\M{R}_j$ is an affine transformation between frames, and
$\M{D}_j$ corresponds to deformation of mesh vertices. Therefore
$\M{M}_i= \sum_{j} \M{D}_j \M{R}_{j}^T$,
where $\M{R}_{j}$ is an affine transformation between frames $0$ and $j$.

The output of this step is sequence $\set{R}=(\M{R}_{1},\dots,\M{R}_{F}$) of
transformation matrices between frame $0$ and all consecutive ones, as well as
a new, transformed data tensor
$\tensor{X}: \forall_i \tensor{X}_{::i}=\tensor{T}_{::i} \M{R}_{i}^T$.

\subsection{Higher Order Singular Value Decomposition}
For the purpose of our compression algorithm, data tensor $\tensor{X}$
containing normalised animation frames is decomposed using HO-SVD. The
resulting Tucker operator
$\tucker{C}{\M{U}^{(1)},\textbf{U}^{(2)},\textbf{U}^{(3)}}$ is passed to
further steps of the algorithm.

\subsection{Compression and reconstruction}
\label{subsection:compression_and_reconstruction}
Vertices of a 3D mesh form $K\times J$ matrix $\M{M}$, where $J=3$. The number
of memory units required to store or transmit an animation of $F$ frames, not
considering a set of edges $\set{G}$, may be expressed as:
\begin{equation*}
S=K \times F \times J \times d_s,
\end{equation*}
where $d_s$ is the size of a single floating-point variable, e.g. $d_s=4$
bytes. HO-SVD allows to reduce the amount of memory required to store an
animation, by decomposing data tensor $\tensor{T}$ and storing only the
truncated Tucker operator
$\tucker{\tilde{C}}{{\tilde{U}^{(1)},\tilde{U}^{(2)},\tilde{U}^{(3)}}}$.
Theoretically there are three compression parameters, corresponding to $J$
dimensions of $\tensor{T}$. However, since the reduction of mode-$2$ components
heavily impacts the quality of the reconstructed mesh, we will only consider
the reduction of $K$ mode-$1$ and $F$ mode-$3$ components. The amount of data
required to store the Tucker operator
$\tucker{C}{\M{U}^{(1)},\textbf{U}^{(2)},\textbf{U}^{(3)}}$ equals
\begin{equation*}
S^{(\mathrm{hosvd})}=(v \times K + J^2 +f \times F+v\times J\times f)\times d_s,
\end{equation*}
where $v$ corresponds to the number of mode-$1$ and $f$ to mode-$3$ components
kept. Therefore
\begin{equation}
\label{equation:cr}
CR^{(\mathrm{hosvd})}=\frac{S^{(hosvd)}}{S}=
\frac{v \times K + J^2 +f \times F+f \times F+v\times J\times f}{K \times F \times J}.
\end{equation}

For visualization of results, space savings ($SS$) will be used in place of
compression rate, defined as:
\begin{equation}
\label{equation:ss}
SS=(1-CR)100\%,
\end{equation}
so $SS=99\%$ denotes only $1\%$ of data remaining after compression.

In addition, we need to store a set of transformation matrices $\set{R}$,
obtained during the first step of the algorithm. Its size is
$S^{\mathrm{(\set{R})}}=12\times F$, and it will be included in our results.

\subsection{HO-SVD compression parameter estimation}
\label{sub:hosvd_parameters}
Application of HO-SVD for 3D mesh compression requires a strategy of
choosing the proportion of preserved components for each
mode, resulting in the required $CR$. Mode-1 components correspond to spatial
information (vertices) and mode-3 to temporal information (frames). If we
denote the number of preserved mode-$1$ components as $v$ and the number of
mode-$3$ components as $f$, $\frac{v}{f}$ is the Vertices-To-Frames ratio
($VTF$).

We estimate $VTF$ by searching for a pair
($v_{\mathrm{min}}$,$f_{\mathrm{min}}$) that gives the lowest reconstruction
error among candidates obtained by using Algorithm
\ref{algorithm:estimte_parameter_space}. This task is time consuming since the
reconstruction must be performed for every pair. Therefore we considered two
simplified solutions for estimating $VTF$, namely:
\begin{itemize}
\item a \emph{diagonal} strategy, where the number of mode-$1$ and mode-$3$
components is similar,
\item an \emph{iterative} solution, where we search for the best pair, assuming
that the function obtained by interpolating values of the reconstruction error
for pairs from Algorithm \ref{algorithm:estimte_parameter_space} is unimodal.
\end{itemize}

The \emph{diagonal} solution assumes that a similar number of components of
each mode is preserved. Therefore, there is no additional cost associated with
finding $VTF$. Intuitively, this strategy may be suboptimal if there is a large
difference in the number of components in each mode (e.g. an animation with
only a couple of frames, but with a large number of vertices). However, we will
show that usually the \emph{diagonal} strategy is sufficient, and its results
are comparable with a more robust approach.

The \emph{iterative} solution results from our observation that for a list of
parameters obtained from Algorithm \ref{algorithm:estimte_parameter_space}, the
distortion of reconstruction performed by truncating the Tucker operator
{$\tucker{C}{{\M{U}^{(1)},\M{U}^{(2)},\M{U}^{(3)}}}$ can usually be
approximated using an unimodal function. Therefore, a minimum can be estimated
with a simple iterative procedure presented as Algorithm
\ref{algorithm:iterative_vtf}.

\begin{algorithm}[!h]
\tcc{$\delta$ is a tolerance margin.} 
\KwIn{$K$, $F$, $\lambda$, $\delta$}
\KwOut{a sequence of ($v$,$f$) pairs}
$\set{P}$=all pairs ($v$,$f$) such that $v\in \{1,\dots,K\}$,$f\in \{1,\dots,F\}$ and $\Psi(v,f)-\lambda\le\delta$\;
\tcc{List is an empty sequence of pairs.} 
\eIf{$V>F$}
{
\For{$v\in\set{P}$}{
List = create sequence of pairs $(v,f)$ such that $|\Psi(v,f)-\lambda|$ is minimal amongst all values of $f$\;}
}
{
\For{$f\in\set{P}$}{
List = create sequence of pairs $(v,f)$ such that $|\Psi(v,f)-\lambda|$ is minimal amongst all values of $v$\;}
}

\eIf{$V>F$}{Sort(List) by $v$}{Sort(List) by $f$}
\Return List
\label{algorithm:estimte_parameter_space}
\caption{A search for a sequence of ($v$,$f$) parameters, that allow to obtain the truncated Tucker operator $\tucker{\tilde{C}}{{\tilde{\M{U}}^{(1)},\tilde{\M{U}}^{(2)},\tilde{\M{U}}^{(3)}}}$ with the desired compression rate $CR$. $K$ is the number of mesh vertices, $F$ is the number of animation frames, and $\lambda$ denotes the desired $CR$. The relation between ($v$,$f$) parameters and $CR$ is described by Eq.~(\ref{equation:cr}) and will be denoted as $\Psi$.}

\end{algorithm}

\begin{algorithm}[!h]
  \SetKwFunction{FindMinimum}{FindMinimum}
  \SetKwProg{minalg}{Function}{}{}
  \minalg{\FindMinimum{List}}{
\KwIn{List: a sequence of ($v$,$f$) pairs}
\KwOut{Index of the best element $i_{\mathrm{min}}$}
\tcc{$s$ is a number of samples.} 
indices = $s$ indices of a uniformly sampled List\;
\tcc{Errors is an empty sequence.} 
\For {$i\in$ indices}
{
\tcc{$\tensor{T}$ is a data tensor}
\tcc{$\tucker{C}{{\M{U}^{(1)},\M{U}^{(2)},\M{U}^{(3)}}}$ is a tucker operator}
\tcc{$\set{R}$ is a sequence of homography matrices}
$\tucker{\tilde{C}}{{\tilde{\M{U}}^{(1)},\tilde{\M{U}}^{(2)},\tilde{\M{U}}^{(3)}}}$ =Truncate(($\tucker{C}{{\M{U}^{(1)},\M{U}^{(2)},\M{U}^{(3)}}}$)\\
$\tilde{\tensor{T}}$ = 
Reconstruct($\tucker{\tilde{C}}{{\tilde{\M{U}}^{(1)}, \tilde{\M{U}}^{(2)},\tilde{\M{U}}^{(3)}}}$, 
            $VTF$, 
            $CR$, 
            $\set{R}$)\;
Append ($\dd(\tensor{T},\tilde{\tensor{T}})$) to Errors\;
}
$i_{\mathrm{min}}$=ArgMin(Errors)\;
\tcc{$I$ is a limit for iterations.} 
\eIf{recursion depth==$I$}
{\Return $i_{\mathrm{min}}$}
{
\Return \FindMinimum(List[indices[$i_{\mathrm{min}}-1$]:indices[$i_{\mathrm{min}}+1$]])
}
}
\label{algorithm:iterative_vtf}
\caption{Estimation of the best pair of parameters ($v$,$f$), that allows reconstruction of $\tensor{T}$ from $\tucker{\tilde{C}}{{\tilde{\M{U}}^{(1)},\tilde{\M{U}}^{(2)},\tilde{\M{U}}^{(3)}}}$ with a minimal error.}
\end{algorithm}

\subsection{Reconstruction quality estimation}
Reconstruction errors were measured by using two standard metrics:
\begin{itemize}
\item Mean Squared Error: $\dd_{\mathrm{MSE}}(\mathbf{v},\mathbf{v}^\prime) = \frac{1}{n}
\sum_{i=1}^{n}(\mathbf{v}^\prime-\mathbf{v})^2$, 
where $\mathbf{v}$ is the original data vector and $\mathbf{v}^\prime$ is its
reconstruction.
\item Hausdorff distance: 
\[\dd_{\mathrm{H}}(\set{A},\set{B})=\max\{ \adjustlimits\sup_{x\in \set{A}} \inf_{y\in \set{B}}
\dd_{\mathrm{e}}(x,y),
\adjustlimits \sup_{y\in \set{A}} \inf_{x\in \set{B}} \dd_{\mathrm{e}}(x,y) \},
\]
where $\set{A}$ is the original, $\set{B}$ -- a reconstructed data set and $\dd_{\mathrm{e}}$ denotes the euclidean distance.
\end{itemize}

Since these metrics may not correspond well with human perception of quality
for 3D objects, an additional metric called Mesh Structural Distortion Measure
(MSDM) described in \cite{Guillaume:2006} was applied. This metric compares two
shapes based on differences of curvature statistics (mean, variance,
covariance) over their corresponding local windows. A global measure between
the two meshes is then defined by the Minkowski sum of the distances over local
windows.

\subsection{Comparison of HO-SVD and PCA application for 3D animation
compression}
In order to verify the performance of HO-SVD, we compared it with another
method of 3D animation compression. Following the idea from \cite{AM2000} we
performed experiments using PCA.

Principal Component Analysis \cite{Jolliffe:2002} may be defined as follows:
 
Let $\M{X}=[\mathbf{x}_1,\mathbf{x}_2\ldots,\mathbf{x}_L]$
be a data matrix, where $\mathbf{x}_i\in\R^{p}$ are data vectors with zero
empirical mean. The associated covariance matrix is given by
$\M{E}=\M{X}\M{X}^T$. 
By performing eigenvalue decomposition of 
$\M{E}=\M{O}\M{D}\M{O}^T$ such that 
eigenvalues $\lambda_i, i=1,..,p$ of $\M{D}$ are ordered in a
descending order
$\lambda_1\geq\lambda_2\geq\ldots\geq\lambda_p>0,$
one obtains the sequence of principal components 
$[\mathbf{o}_{1},\mathbf{o}_{2},\ldots,\mathbf{o}_{p}]$ which are columns of 
$\M{O}$. 
One can form a feature vector $\mathbf{y}$ of dimension $p'\leq p$ by 
calculating 
$\mathbf{y} = [\mathbf{o}_{1},\mathbf{o}_{2},\ldots,\mathbf{o}_{p'}]^T \mathbf{x}.$

In order to apply PCA, tensor $\tensor{T}=t_{i,j,k}\in \R^{F\times J\times K}$
must be unfolded according to Eq. (\ref{equation:unfolding}). Therefore
mode-1 unfolding is performed so the data is flattened row by row to form
matrix $\M{X}_{\tensor{T}}\in \R^{F \times JK}$.

Compression is performed by storing only a limited number of principal
components of $\M{E}$. When reconstructing matrix $\M{X}$, the dimension of
the desired feature vector $p'$ equals the number of principal components
$\mathbf{y} = [\mathbf{o}_{1},\mathbf{o}_{2},\ldots,\mathbf{o}_{p'}]^T \mathbf{x}$ 
used for its calculation and is the only parameter.
The ratio of reduction depends on number $f'$ of the key-frames left. The
compression rate for an animation of a 3D mesh using PCA can be expressed as:
\begin{equation*}
CR^{(pca)}=\frac{(V \times J + F)\times f' \times d_s}{S} 
\end{equation*}

\section{Results}\label{sec:results}
Presentation of results is performed by using a set of well-known
3D animations, summarised in Table \ref{tab:meshes}. \emph{Chicken} and \emph{Gallop} are artificial
sequences of moving animal models. \emph{Collapse} uses the same model as \emph{Gallop} but the applied
deformation is an elastic, non-rigid transformation. \emph{Samba}, \emph{Jumping}, \emph{Bouncing} are motion
capture animations of moving and dancing humans.
\begin{table*}
\begin{center}
\begin{threeparttable}
\begin{tabular}{llccp{1.8in}}
\hline  
       Name             & Referenced as & Vertices & Frames&Description\\ 
\hline  Chicken Crossing\tnote{a} & \emph{Chicken} & $3030$ & $400$ & animation\\ 
  Horse Gallop\tnote{b}     & \emph{Gallop} & $8431$ & $48$  & animation\\
  Horse Collapse   & \emph{Collapse} & $8431$ & $48$  & animation\\  
  Samba\tnote{c}   & \emph{Samba} & $9971$ & $174$ & motion capture sequence\\ 
  Jumping          & \emph{Jumping} & $10002$& $149$ & motion capture sequence\\ 
  Bouncing         & \emph{Bouncing} & $10002$& $174$ & motion capture sequence\\ 
\hline
\end{tabular}
 \begin{tablenotes}
   \item[a] \emph{Chicken} animation was published by Jed Lengyel
   \small{(\url{http://jedwork.com/jed})}
   \item[b] \emph{Gallop} and \emph{Collapse} animations, described in
   \cite{James:2005}, were obtained from the website of Doug L. James and
   Christopher D. Twigg
   \small{(\url{http://graphics.cs.cmu.edu/projects/sma})}.
   \item[c] Motion capture sequences were obtained from the website of Daniel
   Vlasic  \small{(\url{http://people.csail.mit.edu/drdaniel/mesh_animation})}.
 \end{tablenotes}
 \end{threeparttable}
\caption{An overview of animations used for visualization of results.}
\label{tab:meshes}
\end{center}
\end{table*}

Parameter selection for HO-SVD decomposition has an impact on the
reconstruction quality. These parameters are related to the proportion of
preserved components in available tensor modes. 
Fig.~\ref{fig:res:parameter_selection} presents an effect of parameter
selection on MSE reconstruction quality of the \emph{Chicken} animation. Panel
(a) shows how the reconstruction error drops sharply as the number of mode-$1$
and mode-$3$ components grows. For most of the animations, the number of
components that are required for each mode to obtain a reconstruction that is
very similar to the original is small. Panel (b) presents an error value as a
function of possible sets of parameters resulting in the space savings rate of
99\% (so the size of the remaining data tensor is only 1\% of its original
size), as described in subsection \ref{sub:hosvd_parameters}. Annotated
solutions were found by using \emph{diagonal} and \emph{iterative} strategies.
The \emph{diagonal} strategy is fast and from our observations it usually
allows to choose a solution close to the best one. It is therefore a good
``first guess'' choice and can be used if the encoding speed is critical. On
the other hand a solution found by using the \emph{iterative} strategy is more
robust, improving the quality of reconstruction, especially when high
compression ratio is needed.

\begin{figure*}[!h]
\centering
\subfigure[]{\includegraphics[width=0.49\textwidth]{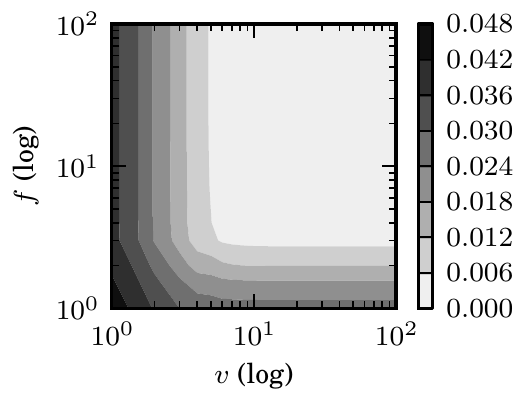}}
\subfigure[]{\includegraphics[width=0.49\textwidth]{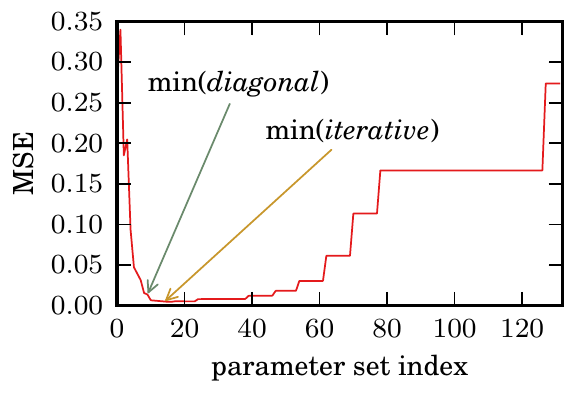}}
\caption{An impact of HO-SVD parameter selection on MSE reconstruction 
for the \emph{Chicken} animation. Panel (a) presents the reconstruction error as a function of the
number of mode-$1$ ($v$) and mode-$3$ ($f$) components, using the logarithmic scale.
Note that the distortion drops sharply with only a few first components. Panel (b)
presents the reconstruction error for possible sets of parameters associated with
$SS=98.8\%$, obtained by using a method described in \ref{sub:hosvd_parameters}.
Annotated sets of best parameters were found by using \emph{diagonal} and
\emph{iterative} strategies. }
\label{fig:res:parameter_selection}
\end{figure*}

Fig.~\ref{fig:parameter_impact} presents $VTF$ ratio for solutions found by using the
\emph{iterative} strategy, as a function of compression ratio. It can be
observed that for high $SS$ values, mode-$3$ components, associated with animated
frames, tend to be more important than mode-$1$ ones, associated with mesh vertices. 

\begin{figure*}[!h]
\centering
\subfigure[]{\includegraphics[width=0.49\textwidth]{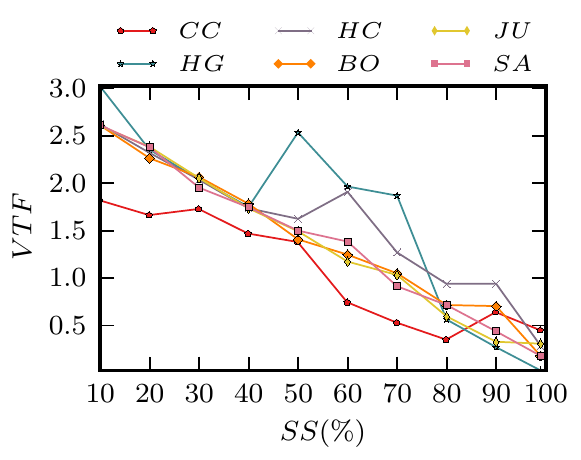}}
\subfigure[]{\includegraphics[width=0.49\textwidth]{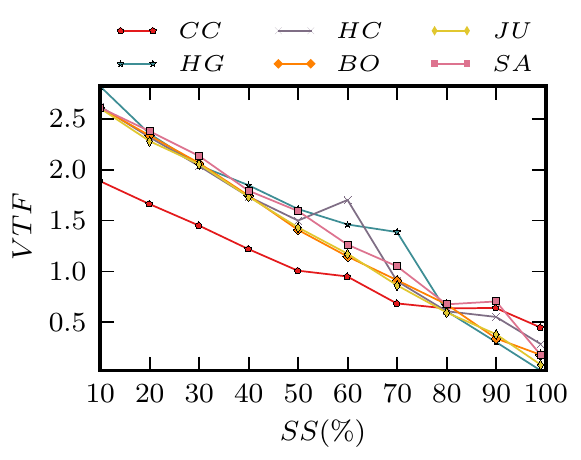}}
\caption{Vertices-to-Frame ratio as a function of $SS$. Experiments were
conducted using the best sets of parameters, selected by using the \emph{iterative}
strategy and (a) Hausdorff, (b) MSE metrics.}
\label{fig:parameter_impact}
\end{figure*}

The distortion introduced by HO-SVD lossy compression, measured with three
quality metrics (MSE, Hausdorff, MSDM) is presented in
Fig.~\ref{fig:res:metrics_ss_function}. The method is robust and introduces low
distortion even for high compression ratio. Observable deformations for
artificial animated meshes (\emph{Chicken}, \emph{Gallop}) are almost
unnoticeable for $SS\sim90\%$ and only minor ones are present for $SS\sim95\%$.
For motion capture sequences (\emph{Samba}, \emph{Jumping}, \emph{Bouncing}),
major deformations are present for $SS\sim95\%$, and only minor ones for
$SS\sim85\%$, with unnoticeable distortions for $SS\sim70\%$. Reconstruction
errors are higher for the \emph{Collapse} mesh, as its animation is hard to
describe using rigid transformations. Major deformations are observable for
$SS\sim90\%$, minor ones are present up to $SS\sim70\%$, and no noticeable
distortions for $SS\sim50\%$ were present. Frames from reconstruction of
animations after HO-SVD compression are presented in Fig.
\ref{fig:visualization_CC} (\emph{Chicken}), Fig. \ref{fig:visualization_HC}
(\emph{Collapse}) and \ref{fig:visualization_SA} (\emph{Samba}). It is worth
noting that despite long computation time, MSDM has advantages over other
metrics: it is bounded with 1.0 as the maximum value (associated with the worst
quality) and it seems to be more sensitive to small, observable distortions of
the mesh surface.

\begin{figure*}[!h]
\centering
\subfigure[]{\includegraphics[width=0.49\textwidth]{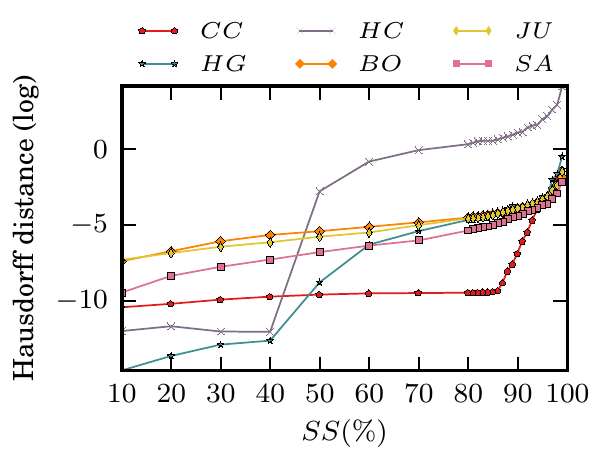}}
\subfigure[]{\includegraphics[width=0.49\textwidth]{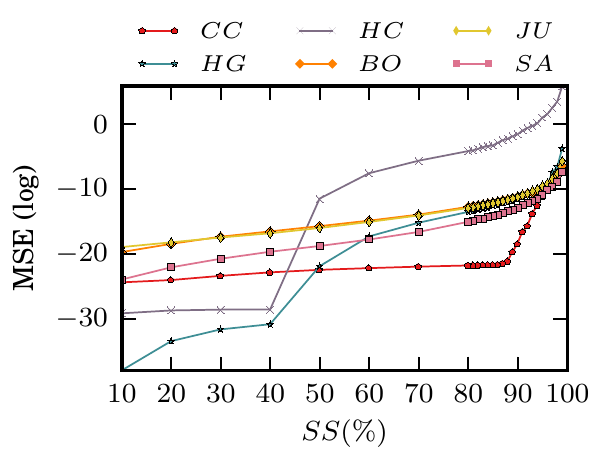}}
\subfigure[]{\includegraphics[width=0.49\textwidth]{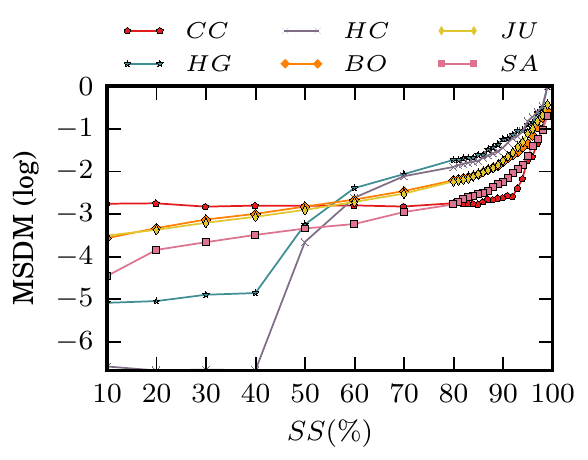}}
\subfigure[]{\includegraphics[width=0.49\textwidth]{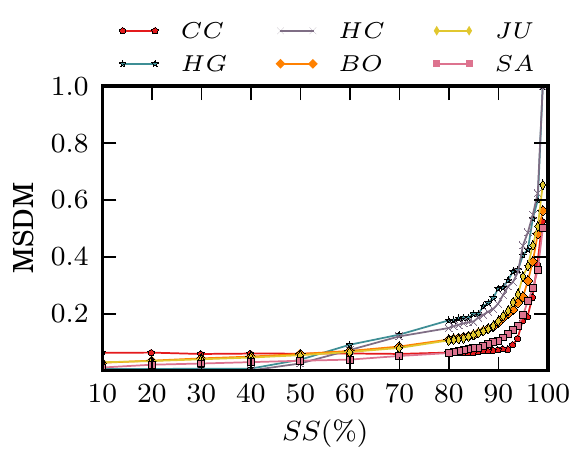}}
\caption{Values of distortion using different 3D quality metrics as a function of
$SS$. Panels (a): Hausdorff distance, (b): MSE (c): MSDM are presented in the
logarithmic scale. Panel (d) presents the results for the MSDM metric.}
\label{fig:res:metrics_ss_function}
\end{figure*}

A comparison of the reconstruction error occurring when using HO-SVD and PCA is presented in
Fig.~\ref{fig:pca_vs_hosvd_artificial} for \emph{Chicken}, \emph{Gallop}, \emph{Collapse} and
Fig.~\ref{fig:pca_vs_hosvd_motion_capture} for \emph{Samba}, \emph{Jumping}, \emph{Bouncing}. HO-SVD compression
gives better result for a majority of animations. Its advantage is visible
especially for motion-capture sequences. Results for \emph{Collapse} show that both methods have problems with describing non-rigid transformations, and their results are similar for high data compression with HO-SVD introducing lower distortion for low compression.

\begin{figure*}[!t]
\centering
	\subfigure[]{\includegraphics[width=0.24\textwidth]{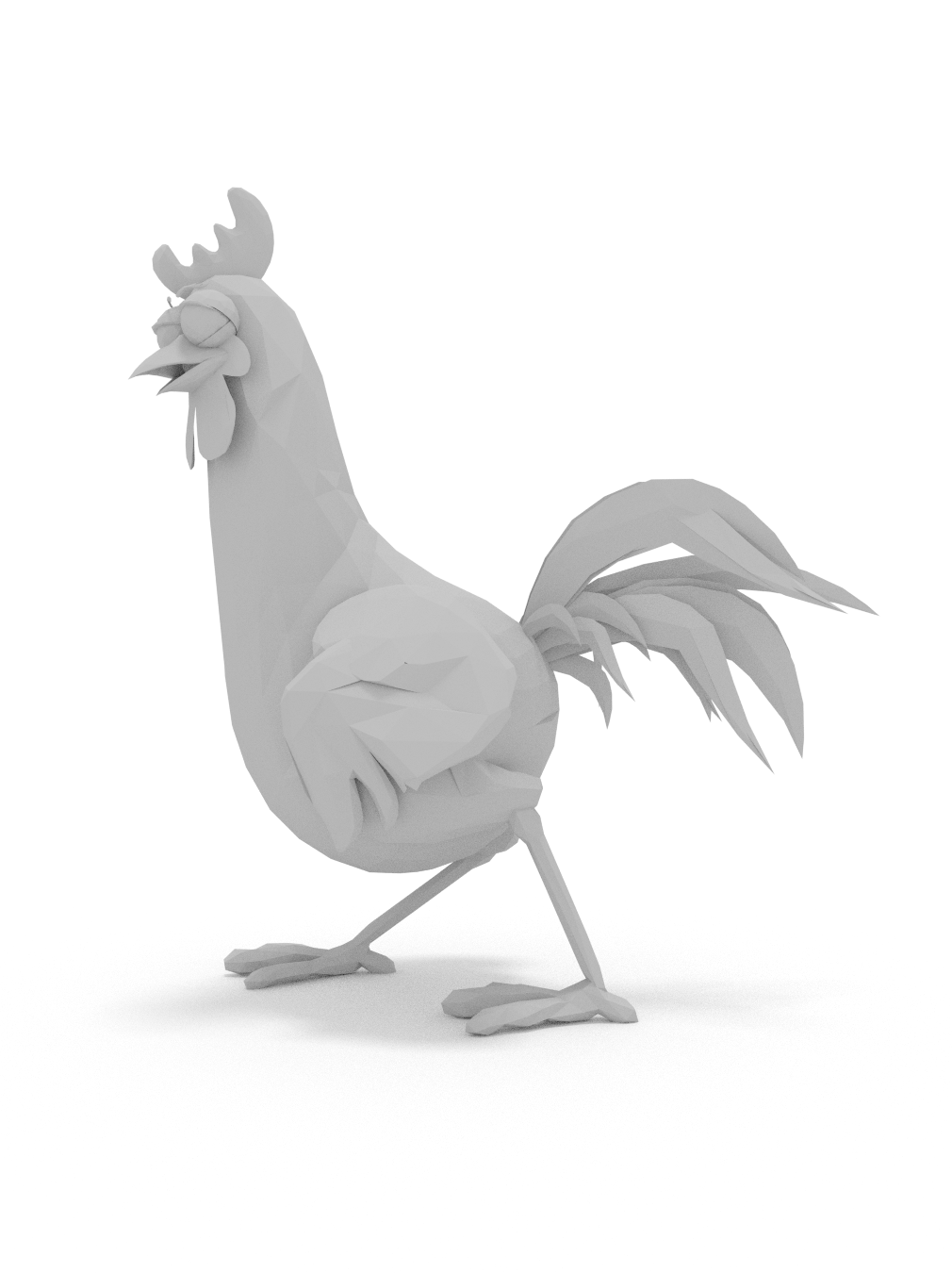}}
	\subfigure[]{\includegraphics[width=0.24\textwidth]{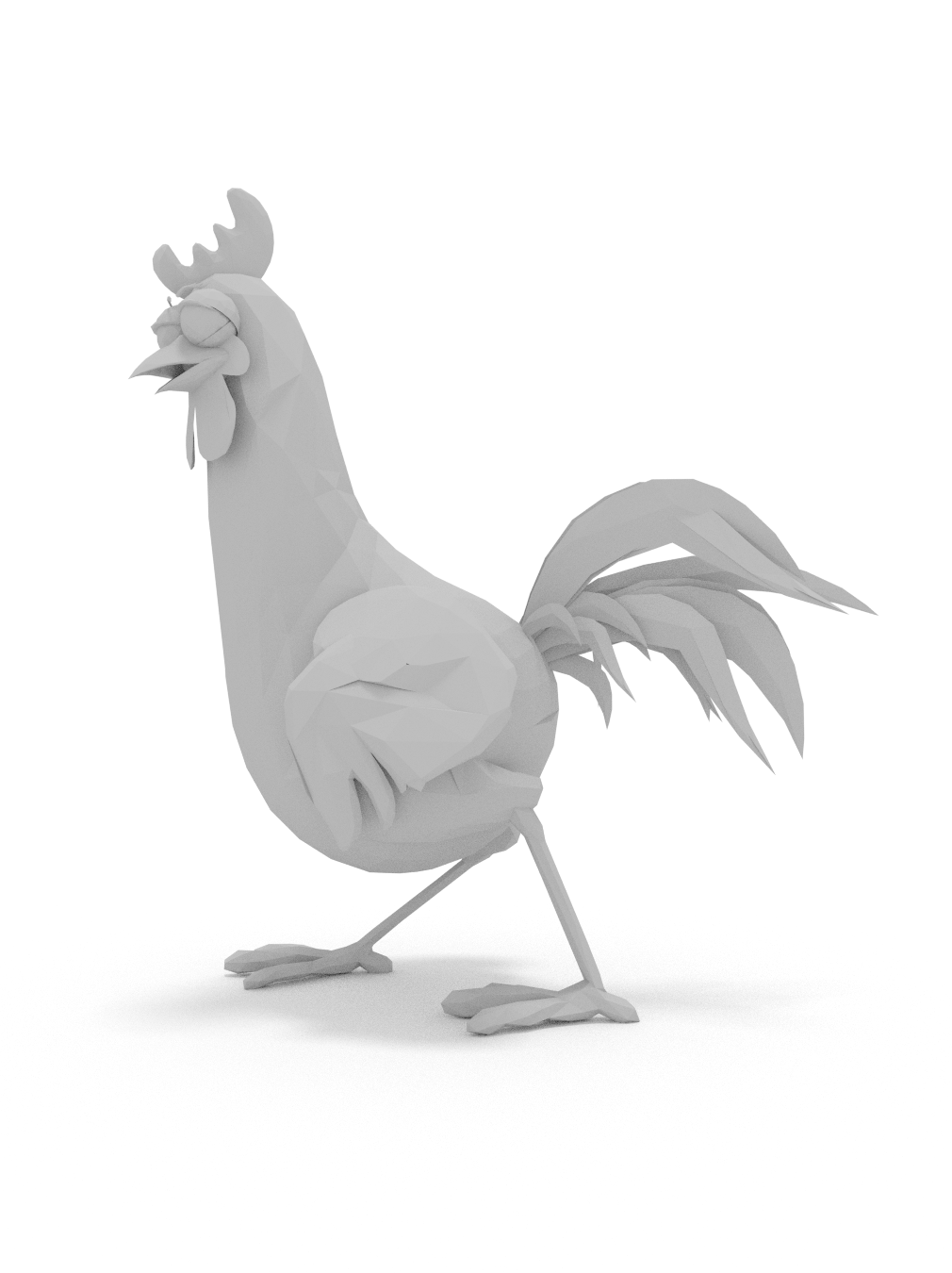}}
	\subfigure[]{\includegraphics[width=0.24\textwidth]{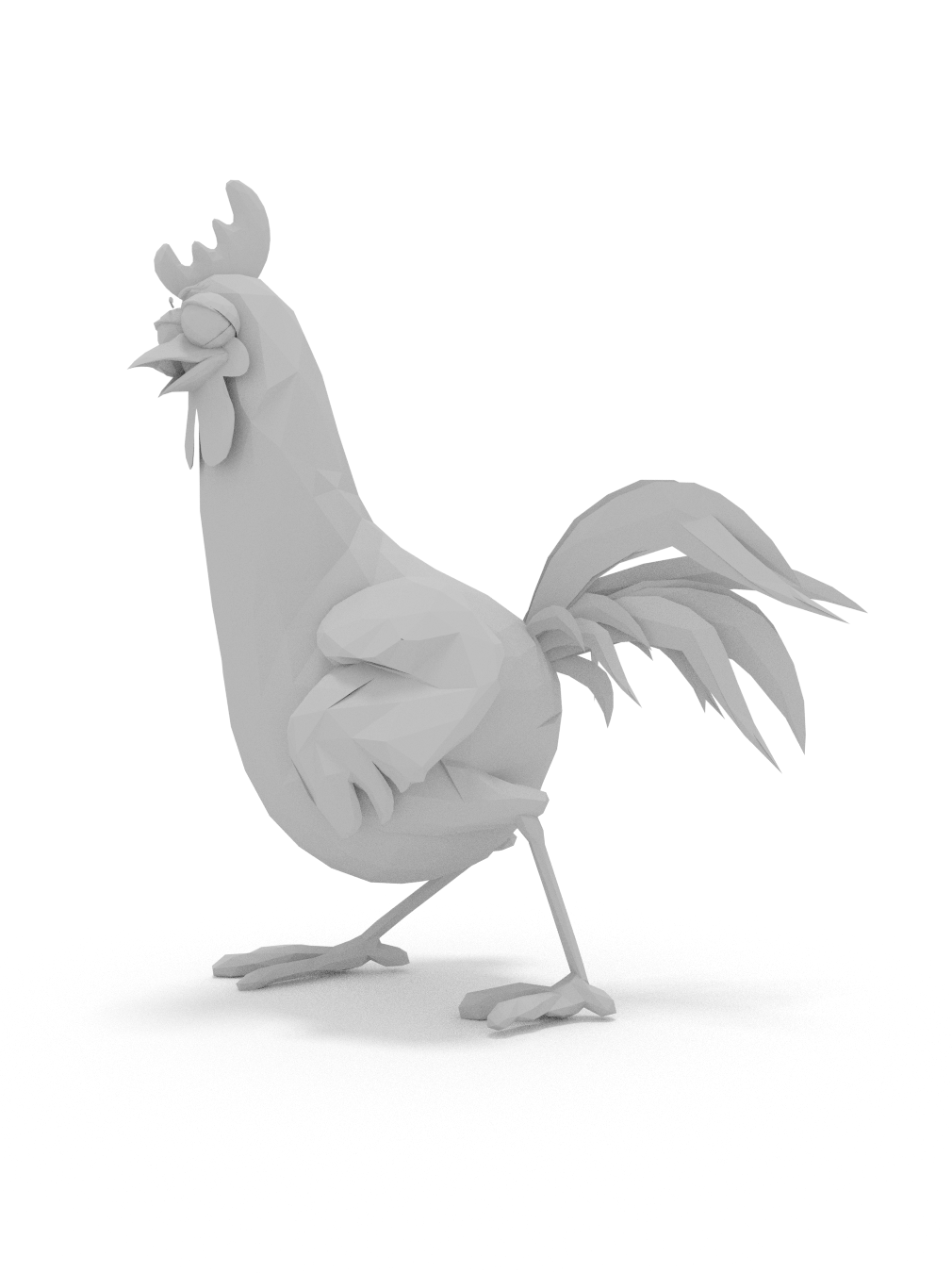}}
	\subfigure[]{\includegraphics[width=0.24\textwidth]{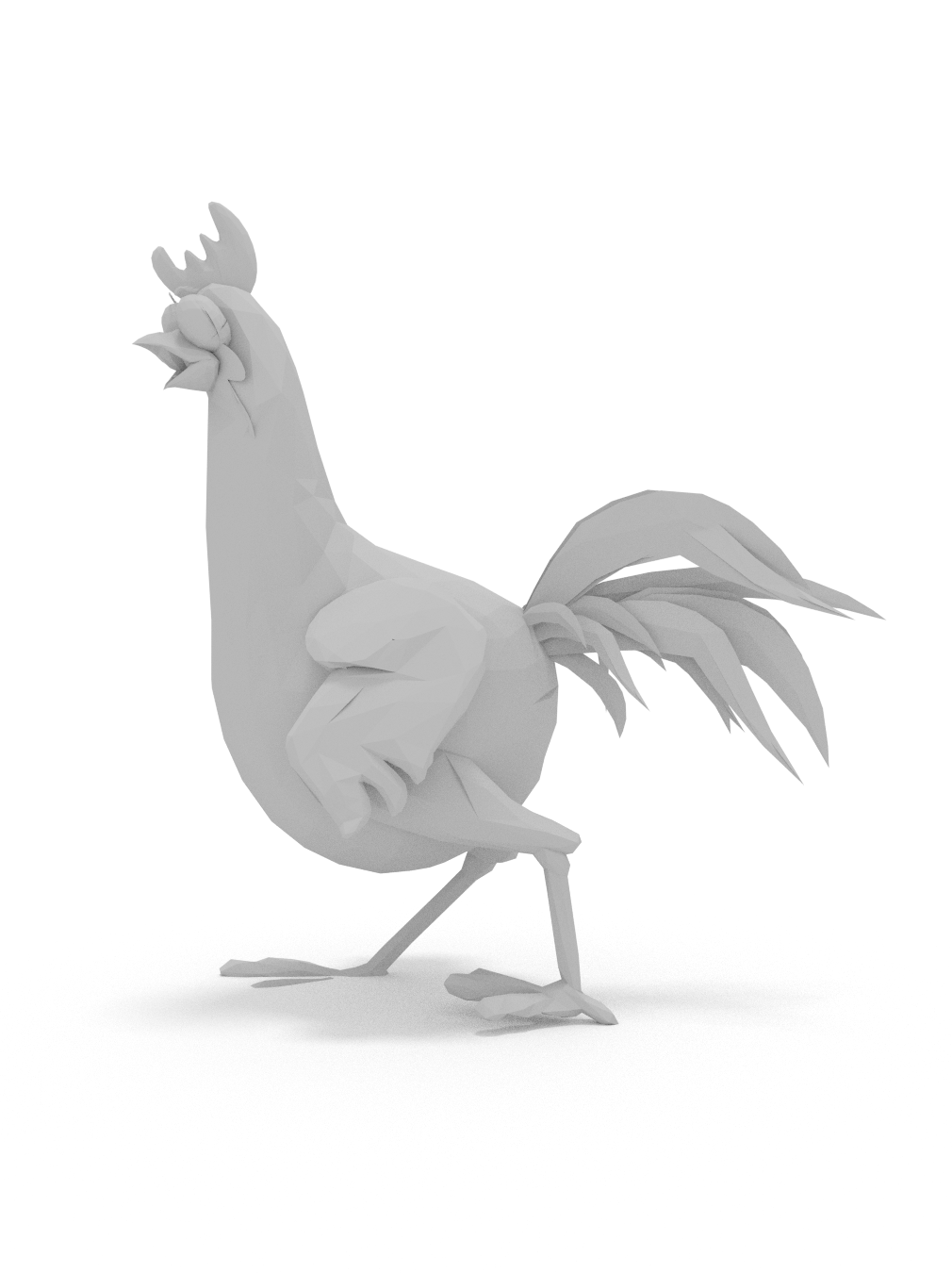}}		
  \caption{Visualization of a reconstructed model for \emph{Chicken}. 
  (a): original, 
  (b): SS=94.8\%,
  (c): SS=97.8\%,
  (d): SS=98.8\%.}
  \label{fig:visualization_CC}
\end{figure*}

\begin{figure*}[!t]
\centering
	\subfigure[]{\includegraphics[width=0.24\textwidth]{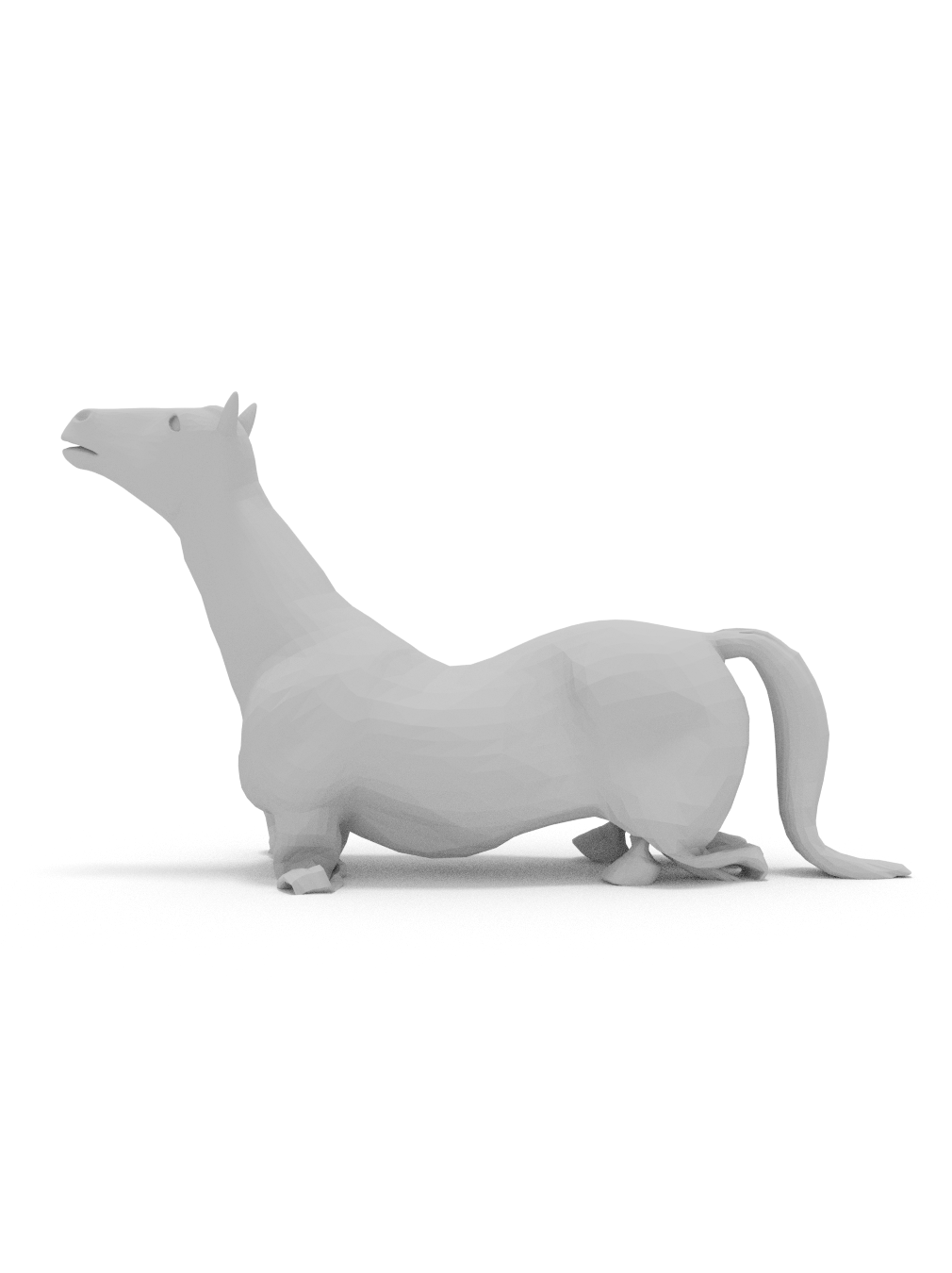}}
	\subfigure[]{\includegraphics[width=0.24\textwidth]{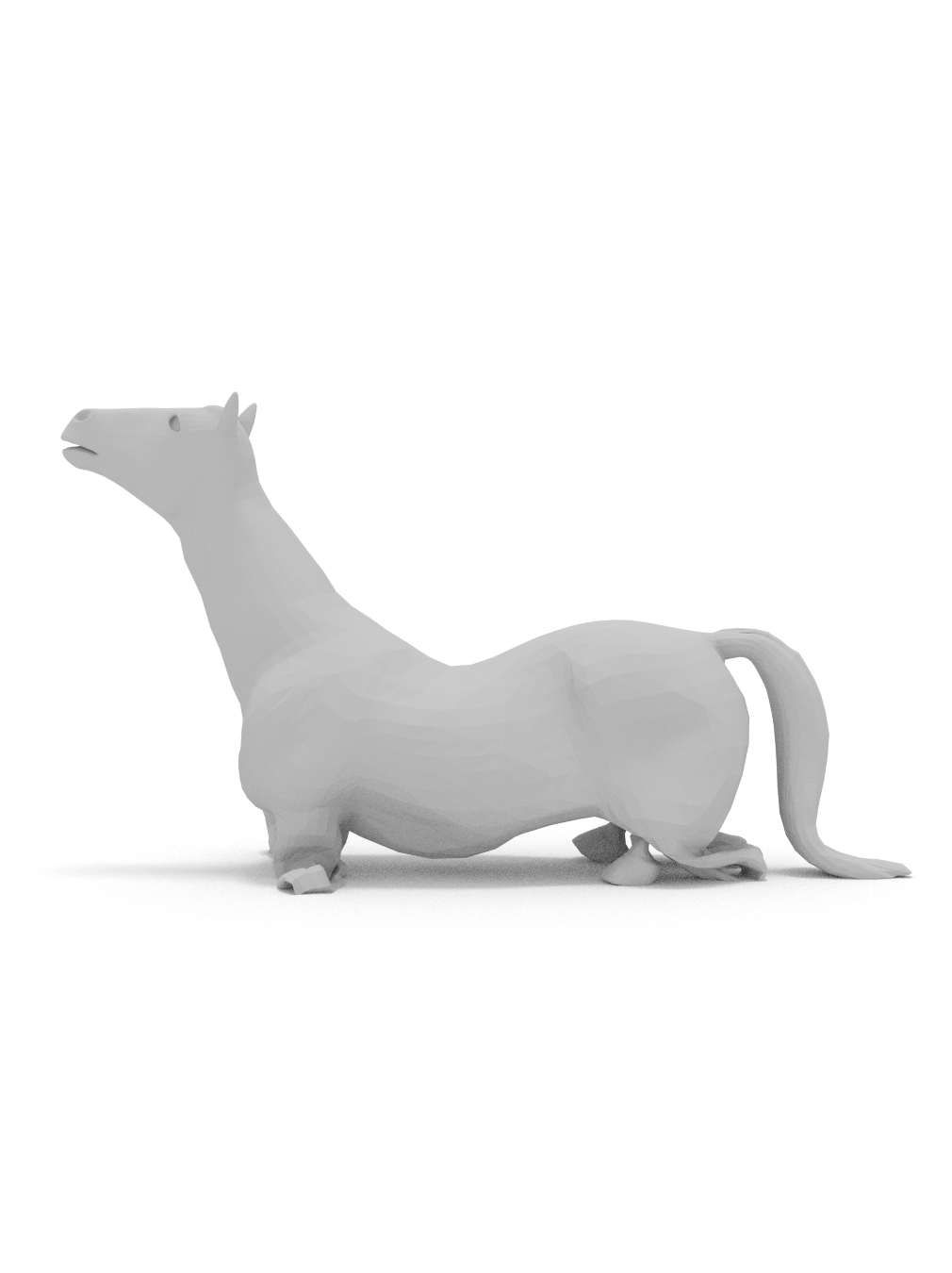}}
	\subfigure[]{\includegraphics[width=0.24\textwidth]{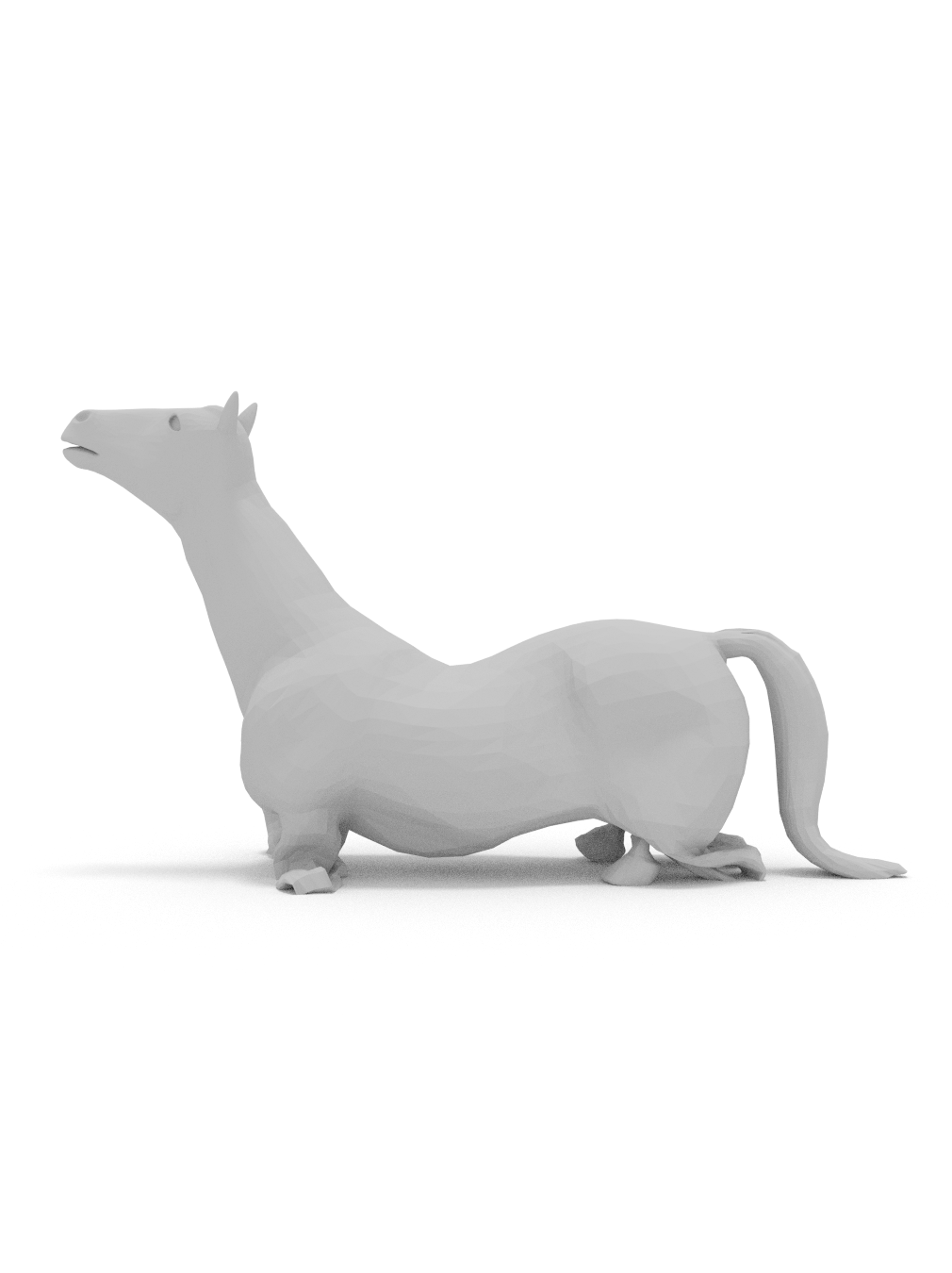}}
	\subfigure[]{\includegraphics[width=0.24\textwidth]{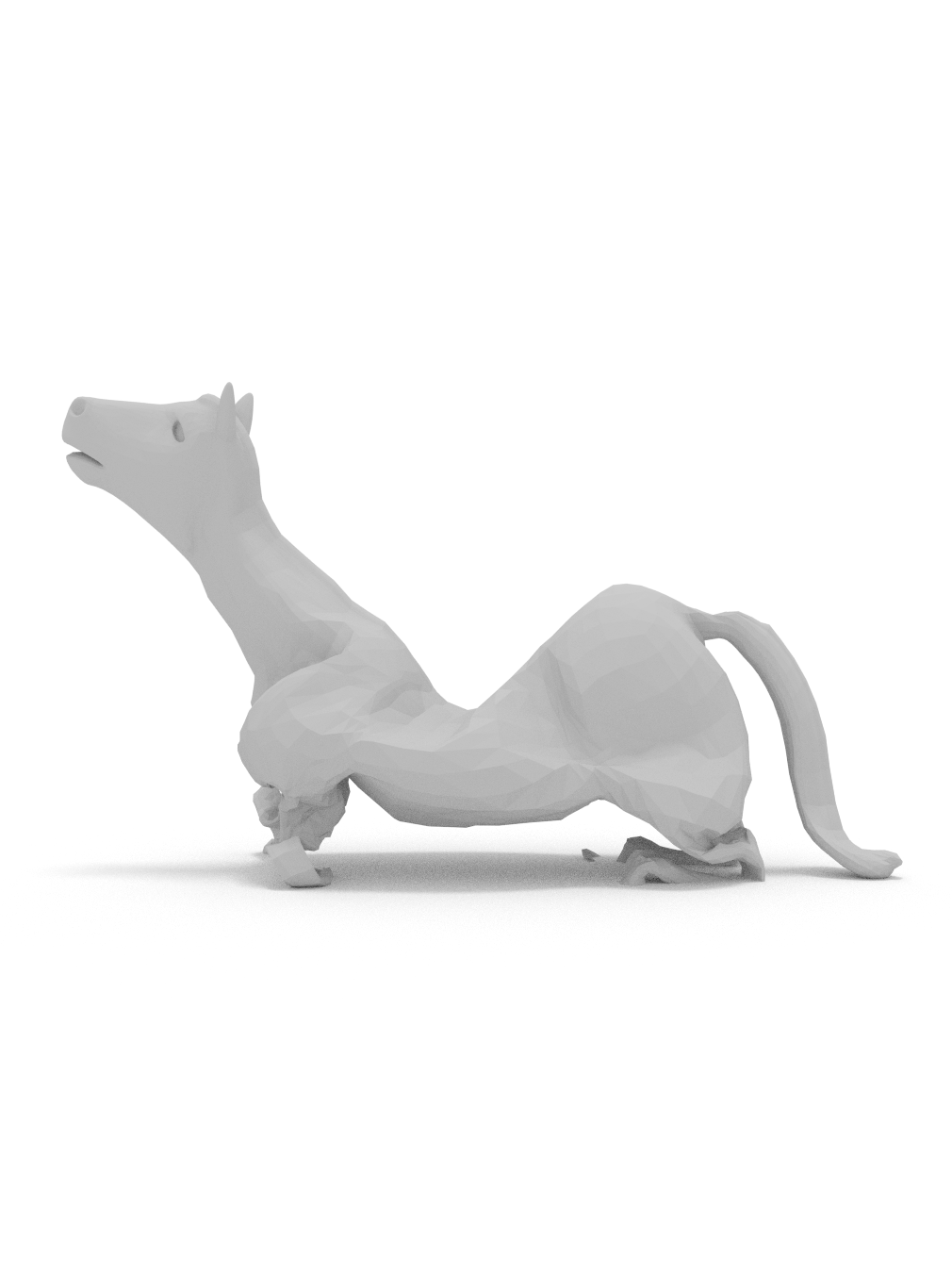}}			
  \caption{Visualization of a reconstructed model for \emph{Collapse}. 
  (a): original, 
  (b): SS=69.9\%, 
  (c): SS=84.9\%, 
  (d): SS=97.9\%.}
  \label{fig:visualization_HC}
\end{figure*}

\begin{figure*}[!t]
\centering
	\subfigure[]{\includegraphics[width=0.24\textwidth]{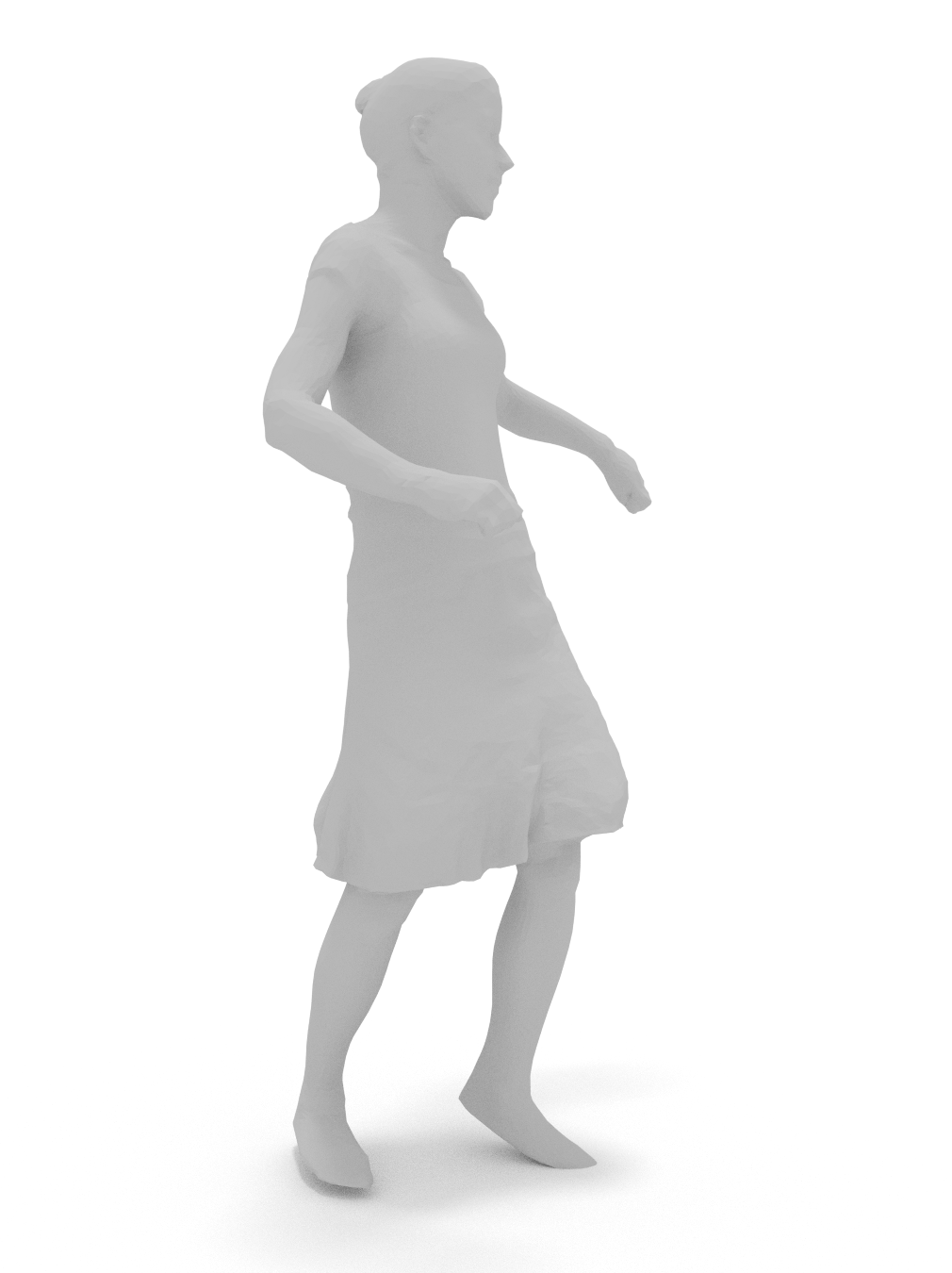}}
	\subfigure[]{\includegraphics[width=0.24\textwidth]{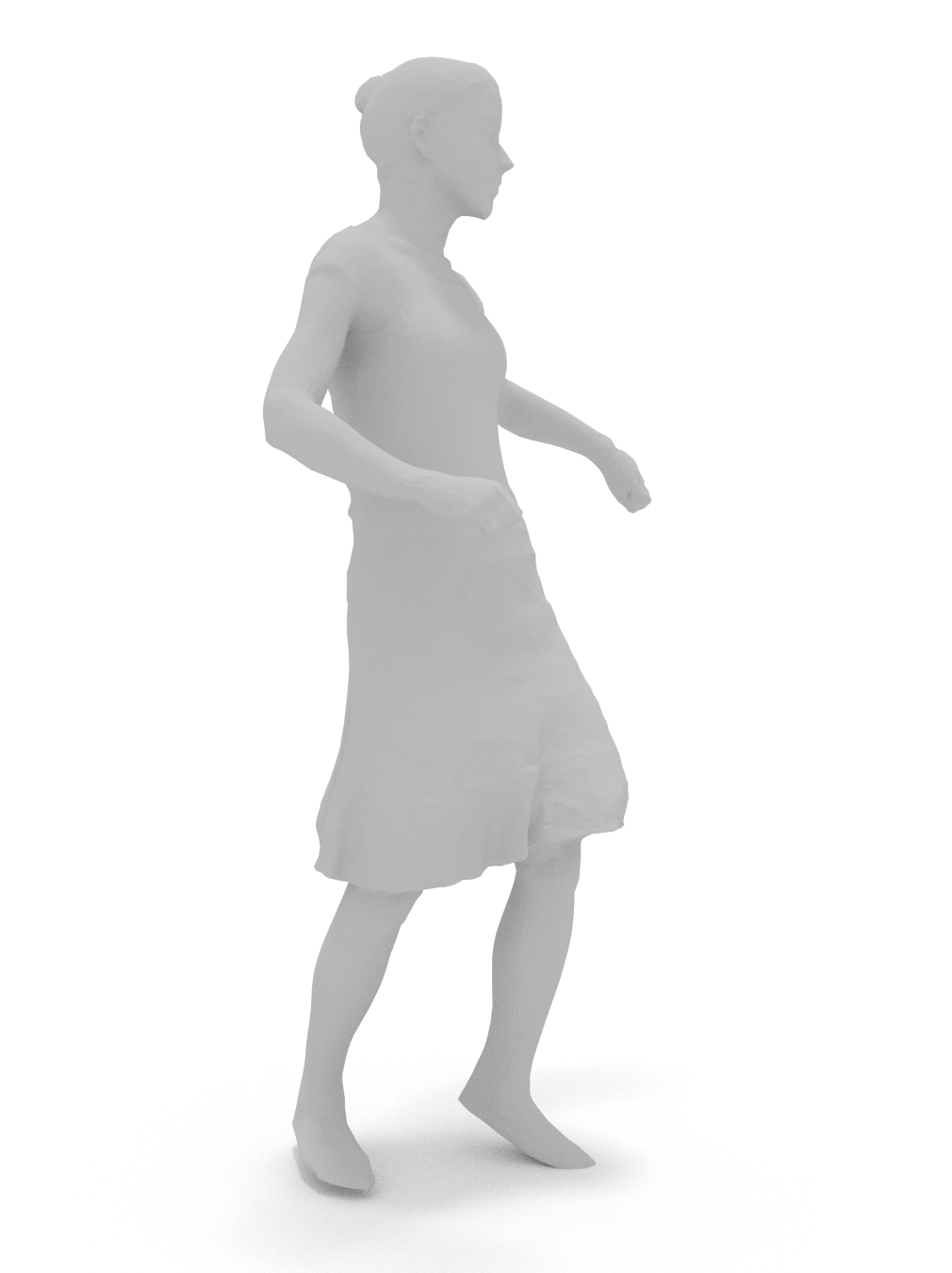}}
	\subfigure[]{\includegraphics[width=0.24\textwidth]{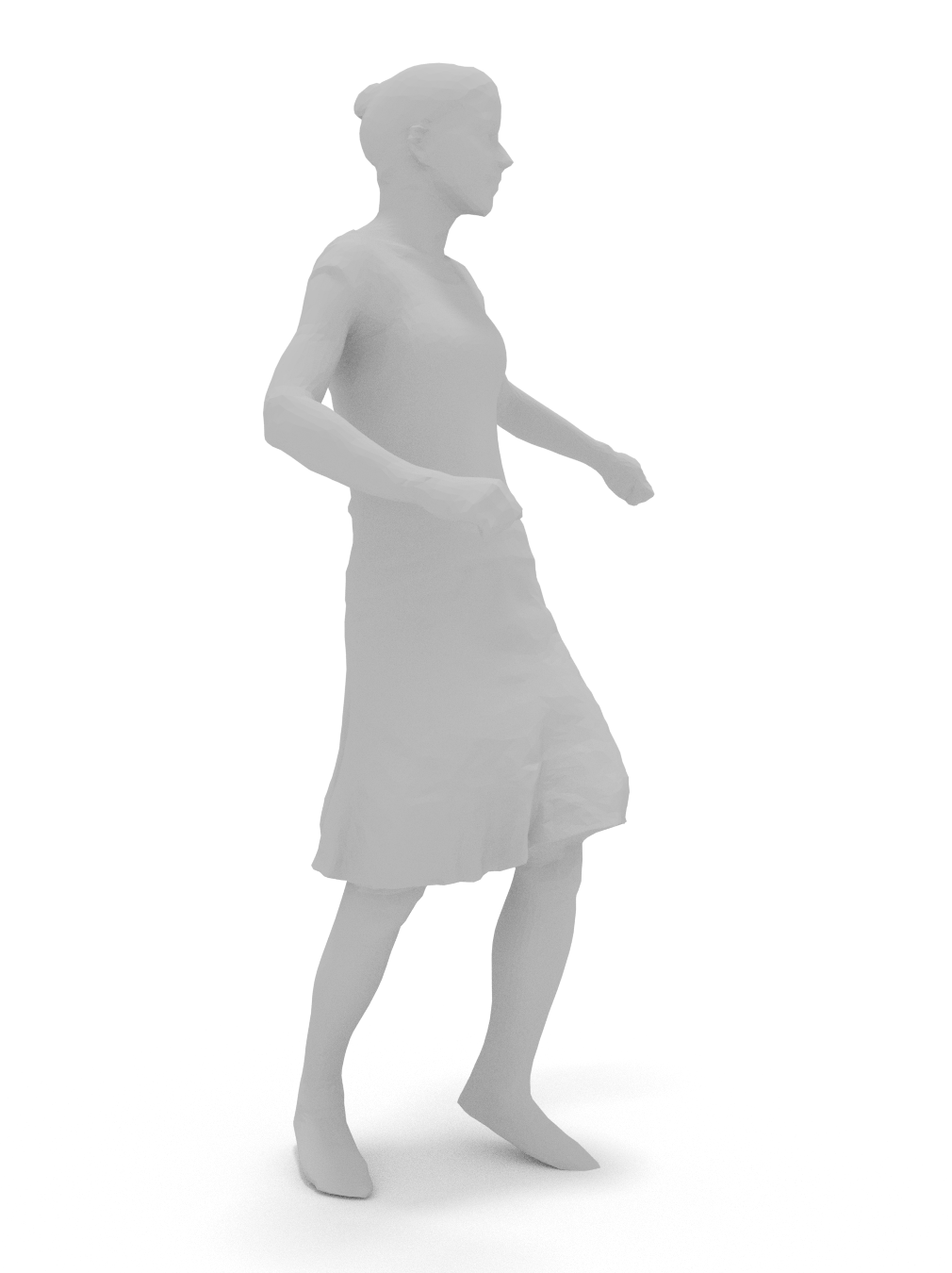}}
	\subfigure[]{\includegraphics[width=0.24\textwidth]{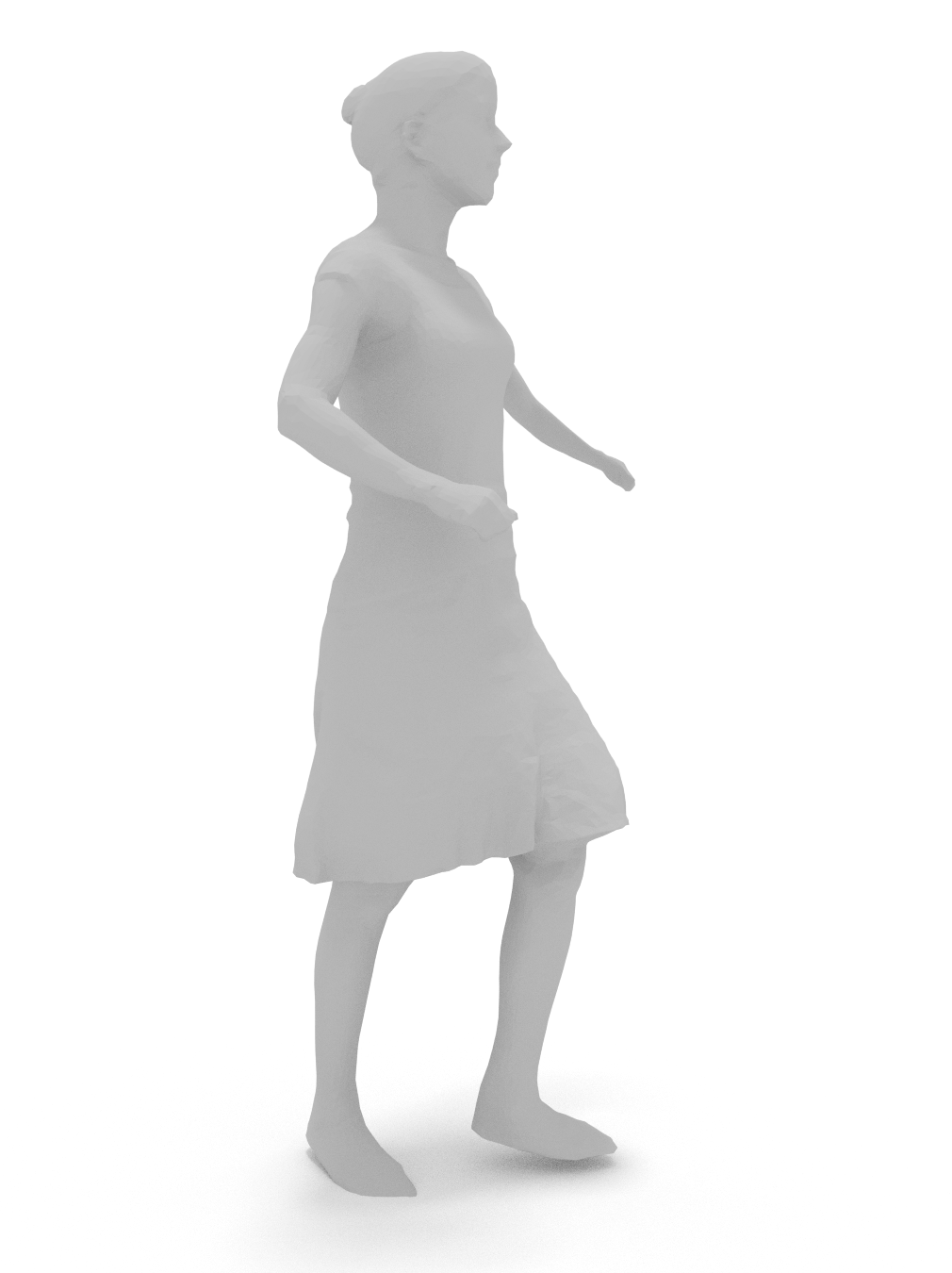}}			
  \caption{Visualization of a reconstructed model for \emph{Samba}. 
  (a): original, 
  (b): SS=89.9\%, 
  (c): SS=94.9\%,
  (b): SS=97.9\%.}
  \label{fig:visualization_SA}
\end{figure*}

\section{Conclusions}\label{sec:conclusions}
Our experiments show that HO-SVD allows to achieve good reconstruction quality
when applied to compression of 3D animations, and usually outperforms the PCA
approach. For most of the animated models and motion-capture sequences,
$SS\sim90\%$ produces a reconstruction similar to the original, especially when
lower level-of-detail is required.

The reconstruction error can be measured by using objective metrics, which
allows reliable control over compression parameters. For determining the
distortion of a reconstructed animation, MSE metric seems to be comparable with
the Hausdorff distance while being fast to compute. On the other hand,
obtaining MSDM is more time-consuming, but the metric is robust and from our
experience it corresponds well with human perception of distortions. Parameters
related to the proportion of preserved components in each mode, after
performing data decomposition, can be found by using the simple heuristic
approach. However, when the speed of coding is critical, the fast strategy for
parameter selection is to choose a similar number of components for each mode,
which usually produces an acceptable solution.

\section*{Acknowledgements}
This work has been partially supported by the Polish Ministry of Science and
Higher Education projects: M. Romaszewski by NN516405137, P. Gawron by
NN516481840, and S. Opozda by NN516482340

\bibliographystyle{eg-alpha}
\bibliography{eigenchickens}

\newlength{\myparbox}
\setlength{\myparbox}{1.4in}
\def\myscale{0.9}
\begin{figure*}[tp]
\hskip-2.0cm
\renewcommand{\arraystretch}{0.5}
\renewcommand{\tabcolsep}{1pt}
\begin{tabular}{c c c c}
& Hausdorff & MSDM & MSE\\
{\begin{sideways}\parbox{\myparbox}{\centering \emph{Chicken}}\end{sideways}}
&\includegraphics[scale=\myscale]{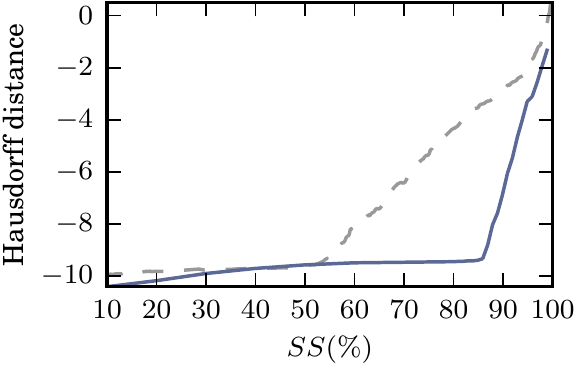}
&\includegraphics[scale=\myscale]{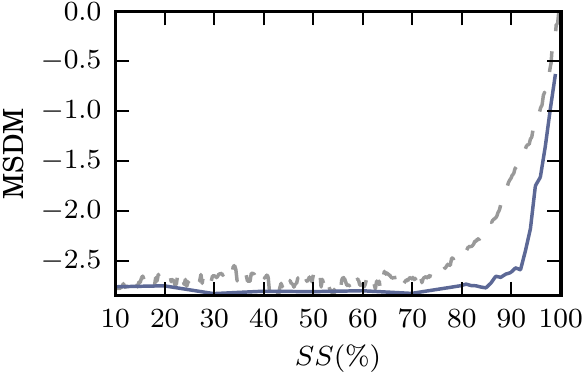}
&\includegraphics[scale=\myscale]{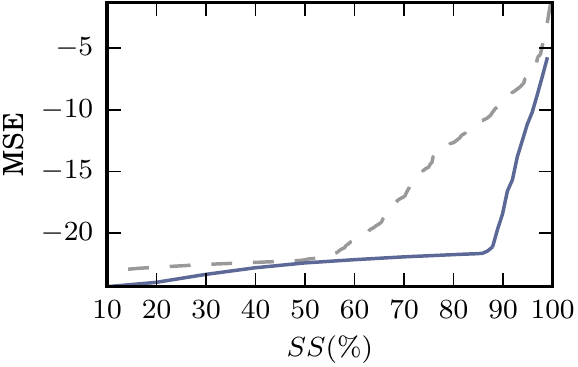}\\ 
{\begin{sideways}\parbox{\myparbox}{\centering \emph{Gallop}}\end{sideways}}
&\includegraphics[scale=\myscale]{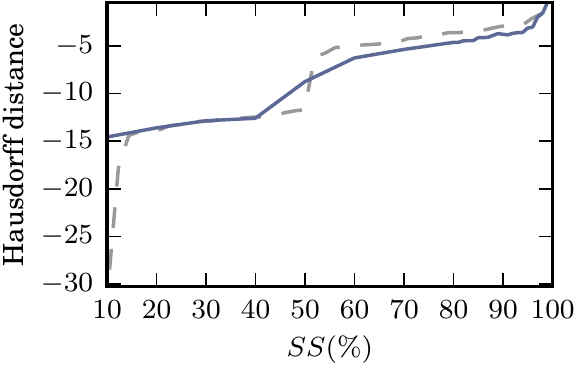}
&\includegraphics[scale=\myscale]{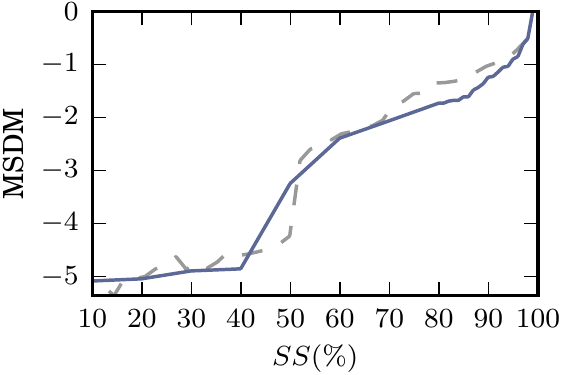}
&\includegraphics[scale=\myscale]{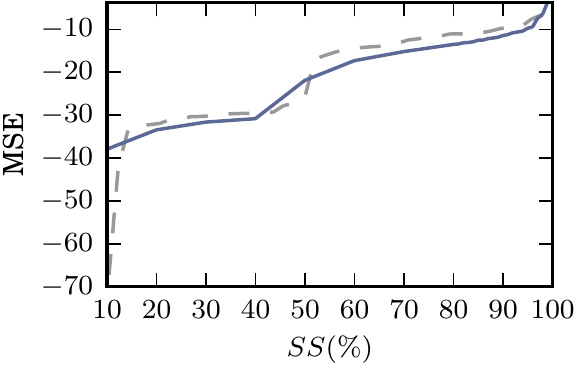}\\ 
{\begin{sideways}\parbox{\myparbox}{\centering \emph{Collapse}}\end{sideways}}
&\includegraphics[scale=\myscale]{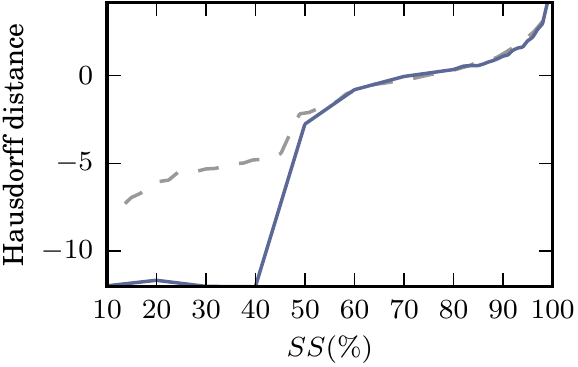}
&\includegraphics[scale=\myscale]{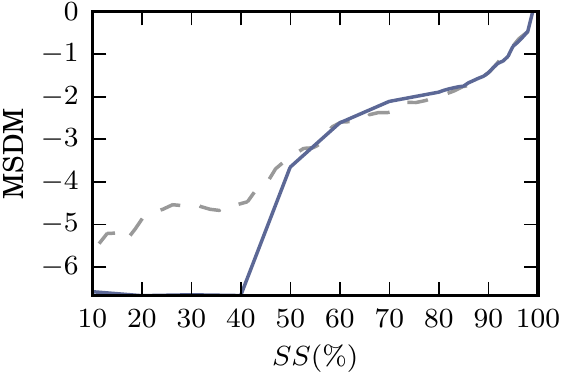}
&\includegraphics[scale=\myscale]{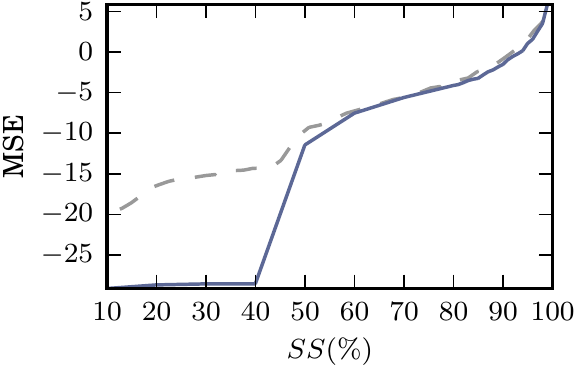}\\ 
\end{tabular}
\caption{A comparison of HO-SVD (solid line) and PCA (dashed line) reconstruction errors for artificial
animations. Distortion is presented in the logarithmic scale as a function of $SS$.
Lower values of distortion indicate higher reconstruction quality.}
\label{fig:pca_vs_hosvd_artificial}
\end{figure*}

\begin{figure*}[tp]
\hskip-2.0cm
\renewcommand{\arraystretch}{0.5}
\renewcommand{\tabcolsep}{1pt}
\begin{tabular}{c c c c}
& Hausdorff & MSDM & MSE\\ 
{\begin{sideways}\parbox{\myparbox}{\centering \emph{Bouncing}}\end{sideways}}
&\includegraphics[scale=\myscale]{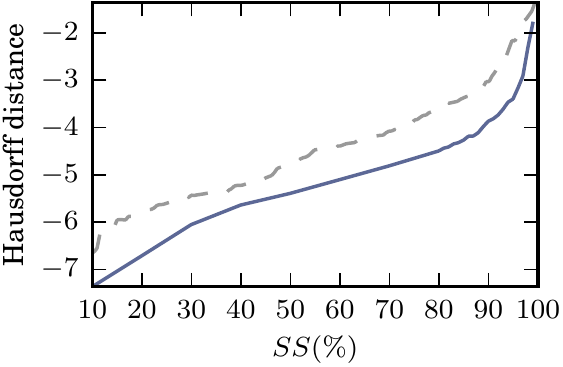}
&\includegraphics[scale=\myscale]{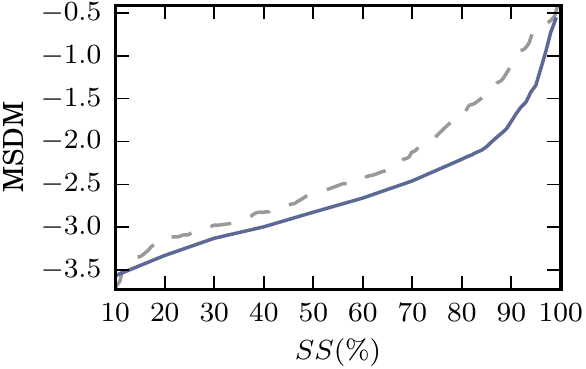}
&\includegraphics[scale=\myscale]{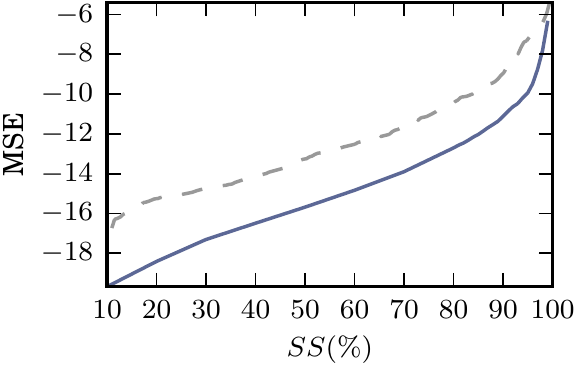}\\ 
{\begin{sideways}\parbox{\myparbox}{\centering \emph{Jumping}}\end{sideways}}
&\includegraphics[scale=\myscale]{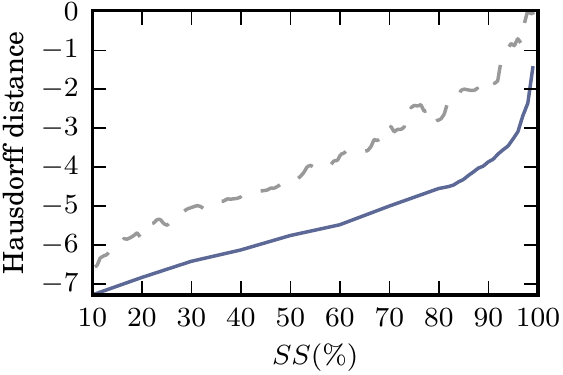}
&\includegraphics[scale=\myscale]{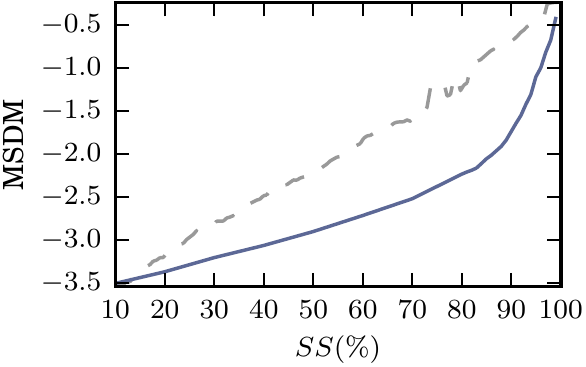}
&\includegraphics[scale=\myscale]{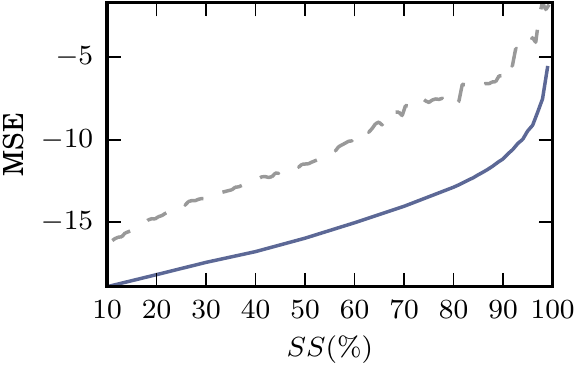}\\ 
{\begin{sideways}\parbox{\myparbox}{\centering \emph{Samba}}\end{sideways}}
&\includegraphics[scale=\myscale]{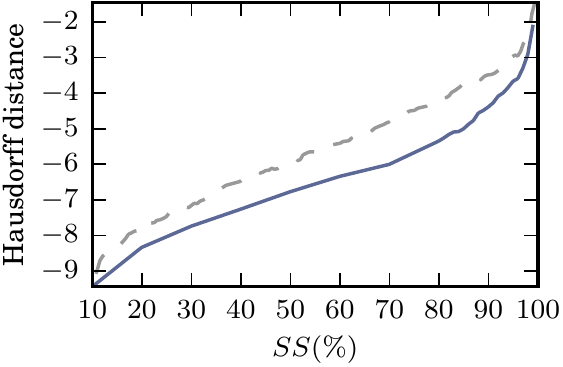}
&\includegraphics[scale=\myscale]{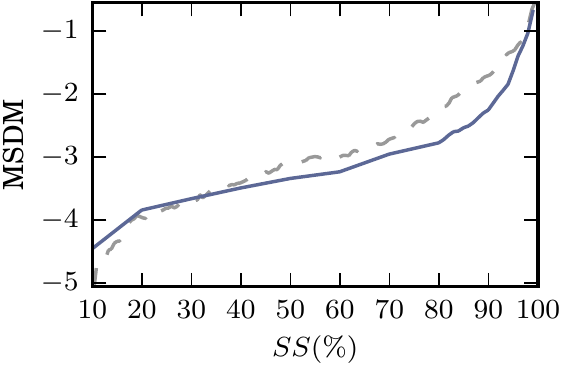}
&\includegraphics[scale=\myscale]{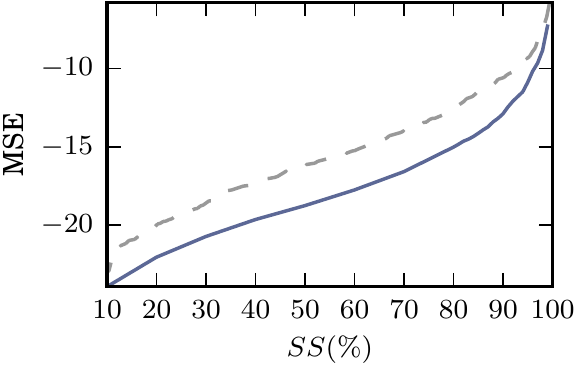}\\ 
\end{tabular}
\caption{A comparison of HO-SVD (solid line) and PCA (dashed line) reconstruction errors for artificial
animations. Distortion is presented in the logarithmic scale as a function of $SS$.
Lower values of distortion indicate higher reconstruction quality.}
\label{fig:pca_vs_hosvd_motion_capture}
\end{figure*}

\end{document}